\documentclass[fleqn,usenatbib]{mnras}

\usepackage[T1]{fontenc}
\usepackage{ae,aecompl}

\usepackage{graphicx}	
\usepackage{amsmath}	
\usepackage{amssymb}	
\usepackage{caption}
\usepackage{pdflscape}

\title[What lies beyond exo-Jupiters?]{High resolution ALMA and HST images of q$^1$~Eri: an asymmetric debris disc with an eccentric Jupiter}

\author[J. B. Lovell et al.]{J. B. Lovell$^{1}$\thanks{E-mail: jl638@cam.ac.uk}, S. Marino$^{1,2}$, M. C. Wyatt$^{1}$, G. M. Kennedy$^{3}$, M. A. MacGregor$^{4}$, \newauthor K. Stapelfeldt$^{5}$, B. Dent$^8$, J. Krist$^5$, L. Matr\`a$^6$, Q. Kral$^7$, O. Pani\'c$^9$, \newauthor T. D. Pearce$^{10}$, D. Wilner$^{11}$\\
$^{1}$Institute of Astronomy, University of Cambridge, Madingley Road, Cambridge, CB3 0HA, UK\\
$^{2}$Jesus College, University of Cambridge, Jesus Lane, Cambridge CB5 8BL, UK \\
$^{3}$Department of Physics, University of Warwick, Coventry CV4 7AL, UK \\
$^{4}$Department of Astrophysical and Planetary Sciences, University of Colorado, 2000 Colorado Avenue, Boulder, CO 80309, USA \\
$^{5}$Jet Propulsion Laboratory, M/S 321-161, 4800 Oak Grove Drive, Pasadena, CA 91109 USA\\
$^6$School of Physics, National University of Ireland Galway, University Road, Galway, Ireland H91 TK33\\
$^7$LESIA, Observatoire de Paris, Universit{\'e} PSL, CNRS, Sorbonne Universit{\'e}, Univ. Paris Diderot, \\ Sorbonne Paris Cit{\'e}, 5 place Jules Janssen, 92195 Meudon, France\\
$^8$ESO, Alonso de C\'{o}rdova 3107, Vitacura, Regi\'{o}n Metropolitana, Chile\\
$^9$School of Physics and Astronomy, University of Leeds, Woodhouse, Leeds LS2 9JS, UK\\
$^{10}$Friedrich-Schiller-Universit{\"a}t Jena, Astrophysikalisches Institut, Schillergaesschen 2-3 07745, Jena, Germany\\
$^{11}$Center for Astrophysics | Harvard \& Smithsonian, 60 Garden St., Cambridge, MA 02138, USA}

\date{Accepted 2021 June 1. Received 2021 May 27; in original form 2021 March 19.}

\pubyear{2020}

\usepackage{newtxtext,newtxmath}

\begin{document}
\label{firstpage}
\pagerange{\pageref{firstpage}--\pageref{lastpage}}
\maketitle

\begin{abstract}
We present \textit{ALMA} 1.3\,mm and 0.86\,mm observations of the nearby (17.34\,pc) F9V star q1 Eri (HD\,10647, HR\,506). 
This system, with age ${\sim}1.4$\,Gyr, hosts a ${\sim}2$\,au radial velocity planet and a debris disc with the highest fractional luminosity of the closest 300 FGK type stars. 
The \textit{ALMA} images, with resolution ${\sim}0\farcs5$, reveal a broad (34{-}134\,au) belt of millimeter emission inclined by $76.7{\pm}1.0$ degrees with maximum brightness at $81.6{\pm}0.5$\,au. 
The images reveal an asymmetry, with higher flux near the southwest ansa, which is also closer to the star.
Scattered light observed with the Hubble Space Telescope is also asymmetric, being more radially extended to the northeast. 
We fit the millimeter emission with parametric models and place constraints on the disc morphology, radius, width, dust mass, and scale height. 
We find the southwest ansa asymmetry is best fitted by an extended clump on the inner edge of the disc, consistent with perturbations from a planet with mass $8\,M_{\oplus} {-} 11\,M_{\rm Jup}$ at ${\sim}60$\,au that may have migrated outwards, similar to Neptune in our Solar System.
If the measured vertical aspect ratio of $h{=}0.04{\pm}0.01$ is due to dynamical interactions in the disc, then this requires perturbers with sizes ${>}1200$\,km. 
We find tentative evidence for an 0.86\,mm excess within 10\,au, $70{\pm}22\, \mu$Jy, that may be due to an inner planetesimal belt. 
We find no evidence for CO gas, but set an upper bound on the CO gas mass of $4{\times}10^{-6}$\,M$_{\oplus}$ ($3\,\sigma$), consistent with cometary abundances in the Solar System.
\end{abstract}

\begin{keywords}
circumstellar matter - planetary systems - planets and satellites: dynamical evolution and stability - techniques: interferometric - stars: individual: HD 10647.
\end{keywords}


\section{Introduction}
\label{sec:intro}
The first detections of exoplanetary systems (systems with planets and/or planetesimal belts) were made a few decades ago \citep{Aumann84, Harper84, Wols92, Mayor95}. 
Many hundreds of systems are now known to possess planetesimal belts and many thousands to possess planets\footnote{\url{http://exoplanet.eu}}. 
Most of these planetesimal belts are cold and reside at 10s of au, making them analogs of the Kuiper Belt in our own Solar System. 
They are inferred to exist from observations of emission from dust which is commonly seen to lie in belts around their parent stars and must have been created in collisions between larger planetesimals. 
Such circumstellar dust and the implied planetesimals collectively form what is known as a star's debris disc \citep{Wyatt08, Hughes18}. 

The number of debris discs is continually growing, and several of these are near enough and sufficiently bright to allow high resolution imaging of the discs' structure and sub-structure with mm/sub-mm instruments such as the Atacama Large Millimetre/sub-Millimetre Array (\textit{ALMA}) \citep[see][and references therein]{Matra18}, and at shorter optical wavelengths, for example with the Hubble Space Telescope (\textit{HST}) \citep[see][]{Apai15}. 
In many cases the observed morphologies are believed to be influenced by the presence of large perturbing bodies (e.g., planets), and many systems have been directly observed with multiple planets \citep[e.g., HR~8799 with 4 planets,][]{Marois08, Su09, Marois10}. 
Disc-planet interactions are expected to produce a wide variety of detectable morphologies such as clumps, radial offsets from their stars, spirals and brightness asymmetries, and characterising these can place important constraints on the architecture and evolution of the entire planetary system \citep{Wyatt99, Wyatt03, Lee16, Faramaz19}. 

The planets that have been observed around other stars fit into a few different classes based on their masses, radii and semi-major axes. 
These include objects such as short period super Earths found during transits, outer giants observed by direct imaging, and exo-Jupiters discovered by radial velocity measurements. 
It might be expected that the formation mechanism of different planets is reflected in the properties of their debris discs, and recent studies have explored such connections for Super Earth systems like 61~Vir \citep{Marino17}, and for systems with known outer giants like HR~8799 and Beta~Pic \citep[see][]{Booth16, Matra17}, and for populations of directly-imaged giant planets \citep[see][]{Meshkat17}. 

Exo-Jupiters have masses and semi-major axes respectively in the ranges $0.1{-}3\,M_{\rm{Jup}}$ and $1{-}5$\,au, are observed around ${\sim}5\%$ of stars, and are commonly found to have eccentric orbits \citep{Chiang13}. 
The origin of their eccentric orbits is understood to arise from early stage instabilities in planetary systems which scatter planets into a broad distribution of eccentricities \citep{Juric08}. 
Planetary system instabilities that excite exo-Jupiter eccentricities have been shown to deplete {\it outer} debris discs (those with radii greater than 10s of au), and in turn these outer discs have been shown to dynamically influence closer in planets \citep{Raymond12, Gomes05, Raymond11}. 
Whilst studies of mutual disc-planet interactions might lead us to conclude that debris belts may be more readily depleted if exo-Jupiters are present, there appears to be neither positive nor negative correlation between the presence of exo-Jupiters in systems with debris discs \citep{Bryden09, MoroMartin15, Yelverton20}. 

q$^1$~Eri (HD~10647, HR~506) is an old (1.4\,Gyr), nearby main sequence F9V star, and is an example of a system with both a bright debris disc and an exo-Jupiter, q$^1$~Eri~b \citep{Marmier13}. 
The q$^1$~Eri system is at a distance of $17.344{\pm}0.014\rm{pc}$ \citep{Gaia16, Gaia18}, and the debris disc has the highest fractional luminosity of the closest 300 sun-like (FGK-type) stars, $f{\geq}10^{-4}$ \citep{Liseau08, Sibthorpe18}. 
The planet q$^1$~Eri\,b has a semi-major axis $a{=}2.03{\pm}0.15$\,au, with a mass $M \sin(i){=}0.93{\pm}0.18\,M_{\rm{J}}$, and a low-moderate eccentricity of $e{=}0.15{\pm}0.08$ \citep{Marmier13}. 

The disc of q$^1$~Eri has been resolved both in the optical (scattered light) with \textit{HST} and in the far-IR (thermal emission) with \textit{Herschel}. 
These studies showed the disc to be highly inclined ($i{>}60^{\circ}$) to the plane of the sky \citep{Stapelfeldt07, Liseau08, Liseau10} with emission concentrated at around ${\sim}100$\,au, although only weak constraints could be placed on the disc's inner edge based on the ${\sim}4''$  (${\sim}70$\,au) resolution of \textit{Herschel}, and given the \textit{HST} coronagraph obscuration inside ${\sim}50$\,au. 
More recent analysis with this \textit{Herschel} data has demonstrated that the disc likely has a radius of $R_{\rm{disc}}{=}81.1^{+1.8}_{-1.3}$\,au, and a broad radial width of $\Delta R_{\rm{disc}}{=}71.1^{+1.9}_{-13.3}$\,au \citep{Marshall21}. 
The \textit{HST} data confirms the disc to be asymmetric, with this more extended in the NE than the SW. 
In addition to the belt beyond 70\,au, the flux distribution indicates that an inner warm component at ${\sim}10$\,au is also present \citep{Kennedy14, Schuppler16}. 
This component is yet to be resolved, but may be in close proximity to the exo-Jupiter q$^1$~Eri\,b.

In this work we present new high-resolution (sub-arcsec) \textit{ALMA} observations of q$^1$~Eri to characterise its asymmetric outer belt and to consider the relationship of that belt to the planetary system architecture, in particular to the known exo-Jupiter. 
These measurements were taken over three epochs, covering two wavelengths; one Cycle 3 (2016) observation in Band~6 (${\sim}1.25\,\rm{mm}$) and two Band~7 observations (${\sim}856\,\mu\rm{m}$) in Cycles 3 (2016) and 5 (2018). 
Combining these data provides a resolution of ${\sim}0.8''$, five times better than that achieved by \textit{Herschel}, sufficient to constrain the inner edge of the debris disc and observe the previously unresolved inner regions. 
In addition we present 2006 \textit{HST} data, as included in \citet{Stapelfeldt07}, to complement our analysis of the mm/sub-mm observations. 

We provide an overview of all our observational data sets in $\S$\ref{sec:ALMAObsFull} and discuss our initial observational analysis of these in $\S$\ref{sec:obsAnalysis}. 
We then present our methodology to model this system in $\S$\ref{sec:modelling}, give a discussion of our findings and future work in $\S$\ref{sec:discussion} and summarise our key conclusions in $\S$\ref{sec:conclusions}.

\section{Observations}
\label{sec:ALMAObsFull}
\begin{figure}
    \includegraphics[width=0.5\textwidth]{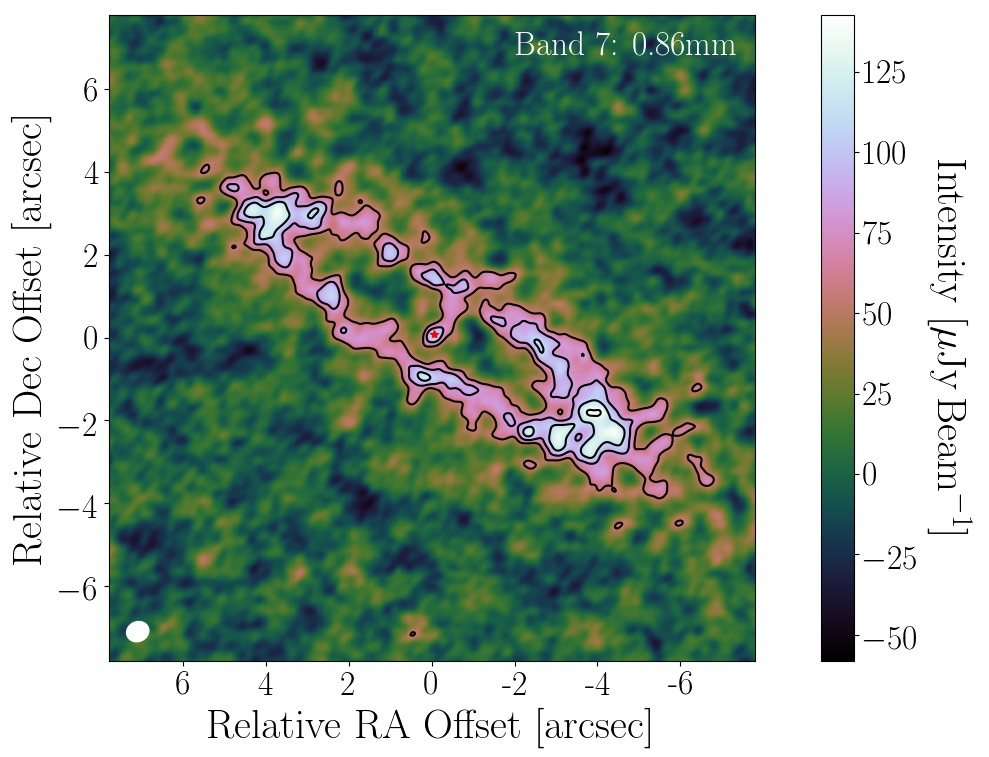}
    \includegraphics[width=0.5\textwidth]{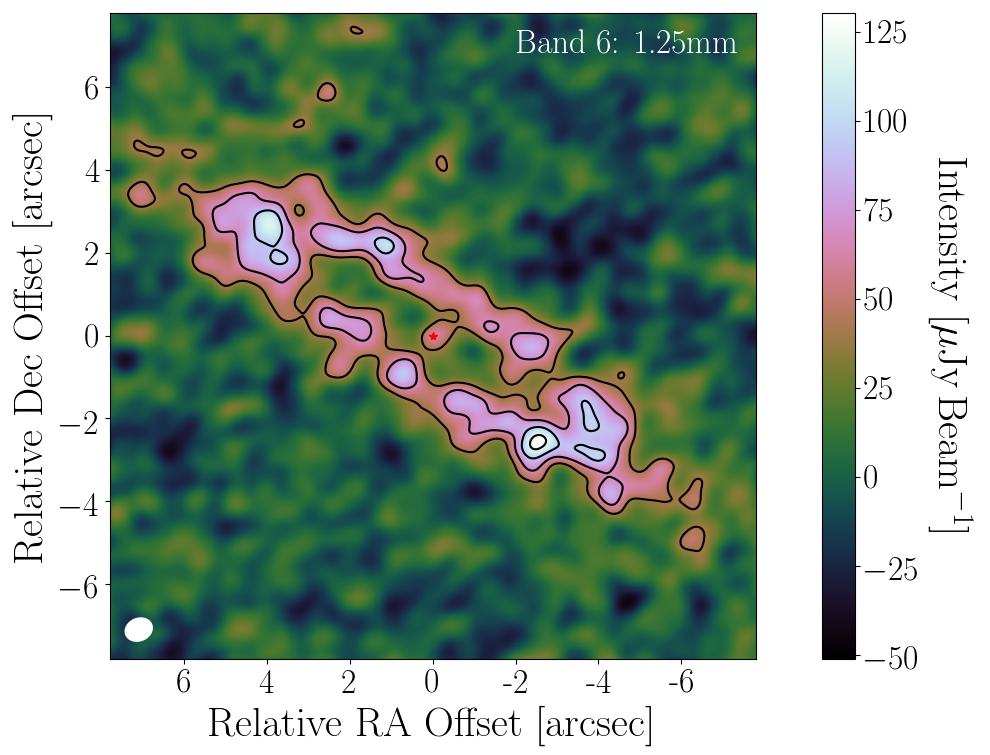}
    \caption{Non-primary beam corrected \textit{ALMA} images of q$^1$~Eri in Band~7 (0.86\,mm, top) and Band~6 (1.25\,mm, bottom), with the synthesised beams shown in the lower left in white, and the stellar location marked with a red star (both in the coordinate centres). In both North is up, East is left. Unresolved emission is detected in the centre of these figures deemed to be mostly from the star. Top: A cleaned Briggs weighted (robust=1.0) image for the combined Band~7 2016 and 2018 data sets, with contours of +4, +6, +8 and +10$\,\sigma$ significance. The beam is $0.55 {\times} 0.47''$, $\rm{PA_{\rm{beam}}}{=}-66.9^{\circ}$, and the image rms is $\,\sigma{=}13.9\,\mu$Jy\,beam$^{-1}$. Bottom: A cleaned natural weighted ${\sim}0.4''$ uv-tapered image in Band~6, with contours of +3, +5, +7 and +9$\,\sigma$ significance. The beam is $0.67 {\times} 0.54''$, $\rm{PA_{\rm{beam}}}{=}-67.6^{\circ}$, and the image rms is $\,\sigma{=}13.3\,\mu$Jy\,beam$^{-1}$.}
    \label{fig:B6B7Image}
\end{figure}

\subsection{Submillimetre ALMA Observations}
\label{sec:ALMAObs}
\subsubsection{ALMA Band~6 Observations}
q$^{1}$~Eri was observed for ${\sim}80$ minutes (on source) in two scheduling blocks with \textit{ALMA} in Band~6 during Cycle 3 as shown in the top panel of Table~\ref{tab:band7Obs}, using 41 antennas with minimum and maximum baselines ranging from 15.1 to 772.8\,m as part of project 2015.1.00307.S (PI: David Wilner). 
The correlator had 3 spectral windows centred on frequencies of $232.490$, $244.989$, and $246.989$\,GHz, for continuum observations with a bandwidth of $2.000$\,GHz and channel widths of $15.625$\,MHz. 
Also set up was a spectral window centred on a topocentric frequency of $230.536$\,GHz with a bandwidth of $1.875$\,GHz, and channel widths of $488.281$\,kHz, for CO J=2-1 spectral line observations. 
The visibility data set was calibrated using the CASA software version 5.1.1-5 with the standard pipeline provided by the \textit{ALMA} Observatory. 
Additional data flagging was performed on the Band~6 data to mitigate issues introduced by several poorly performing antennas. 
The $\rm{plotms}$ task in $\rm{CASA}$ was used to examine the visibility amplitudes as a function of time and uv-distance. 
Outlying points were flagged using the $\rm{flagdata}$ task. 
Table~\ref{tab:band7Obs} shows a summary of the observational setup for \textit{ALMA} data collection. 
Continuum imaging was conducted using the CASA $\rm{tclean}$ algorithm with natural weighting (to enhance S/N), shown in Figure~\ref{fig:B6B7Image} (bottom) in which the disc is clearly detected. 
The synthesised beam size in this image is $0.67 \times 0.54''$ ($\rm{PA}_{\rm{Beam}}{=}69.7^{\circ}$), which at 17.34\,pc corresponds to a physical size of $11.6\times 9.4$\,au.

\subsubsection{ALMA Band~7 Observations}
\label{sec:ALMAB7Obs}
q$^1$~Eri was observed for ${\sim}177$ minutes (on source) by \textit{ALMA} in Band~7 over four scheduling blocks in Cycles 3 and 5 as shown in the middle and lower panels of Table~\ref{tab:band7Obs}, with baselines ranging from 15.1 to 867.2m (2016) and 15.0 to 313.7m (2018). 
In 2016 (project 2015.1.01260.S, PI: Mark Wyatt) using 33 antennas, the correlator had 3 spectral windows centred on frequencies of 347.817, 335.775, and 333.817\,GHz each with a bandwidth of 2.0\,GHz and channel widths of 15.625\,MHz (a total of 128 channels each, for continuum observations). 
Also set up was a spectral window centred on a topocentric frequency of 345.817\,GHz with a bandwidth of 1.875\,GHz and spectral channel widths of 488.281\,kHz (a total of 3840 channels), for CO J=3-2 spectral line observations. 
In 2018 (project 2017.1.00167.S, PI: Mark Wyatt) using 46 antennas (in observation block 1) and 43 antennas (in observation block 2), the same correlator set up (3 continuum, 1 spectral line) and respective channel widths were used, however the central frequencies for the continuum observations were 347.683, 335.683, and 333.788\,GHz, and the spectral line observation central frequency (topocentric) was 345.787\,GHz (still covering the line transition frequency). 

The 2016 Band~7 visibility data sets were calibrated using the CASA software version 4.7.2 with the standard pipeline provided by the \textit{ALMA} Observatory, whilst the 2018 Band~7 visibility data sets were calibrated using the CASA software version 5.1.1-5 (also with the standard pipeline provided by the \textit{ALMA} Observatory). 
Both Band~7 data sets were time averaged to 10-second widths, channel averaged into 4 channels per spectral window, and the CASA $\rm{statwt}$ task was run to estimate the visibility weights based on the measured variance. 
We note here that the Band~7 data in 2016 got a ``semi pass" rating for its quality assurance, since a smaller synthesised beam was obtained in 2016 than requested. 

\begin{table}
    \centering
    \caption{\textit{ALMA} observational setup, over the 6 different epochs. The horizontal lines separate the (top) Band~6, (middle) 2016 Band~7 and (bottom) 2018 Band~7 observation sets. All times represent time on source.}
    \begin{tabular}{c|c|c|c|c|c|c}
         \hline
         \hline
         D.M.Y & Time & Flux & Bandpass & Phase \\
         & [mins] & Calibrator & Calibrator & Calibrator \\
         \hline
         26.05.16&39:58&Pallas&J2258-2758&J0210-5101\\
         02.06.16&39:58&Pallas&J2258-2758&J0210-5101\\
         \hline
         12.07.16&48:34&Ceres&J2258-2758&J0210-5101\\
         12.07.16&48:34&Ceres&J0538-4405&J0210-5101\\
         \hline
         27.06.18&39:58&J2258-2758&J2258-2758&J0124-5113\\
         07.07.18&39:57&J0159-4546&J0159-4546&J0210-5101\\
    \hline
    \end{tabular}
    \label{tab:band7Obs}
\end{table}

To improve the S/N and resolution of the Band~7 data, we combined the two Band~7 data sets. 
Between the 2016 and 2018 observations the proper motion of q$^{1}$~Eri, $\mu_{\rm{RA,Dec}}{=}(165.83{\pm}0.10,-105.52{\pm}0.10)$\,$\rm{mas}\,\rm{yr}^{-1}$, resulted in its position shifting by ${\sim}0.4''$ \citep{Gaia16, Gaia18}.
Since the \textit{ALMA} observation phase centres were not perfectly aligned with the expected stellar position from \textit{Gaia} DR2, we realigned to these coordinates using the CASA task $\rm{fixvis}$. 
We then used the CASA task $\rm{fixplanets}$ to shift the coordinates of the 2018 observations to coincide with the 2016 measurement set phase centre (see Table \ref{tab:posGaia} for these expected DR2 coordinates). 
The two resulting measurement set epochs were then combined with the CASA task $\rm{concat}$. 
We note that correcting for the stellar proper motion in this way would result in background sources (if present) being smeared along the proper motion direction, however we find no evidence of any point sources which appear significantly elongated (i.e., the emission is consistent with a point source at the location of the star). 
Fig.~\ref{fig:B6B7Image} (top) shows the resulting Band~7 cleaned image of q$^{1}$~Eri with the CASA $\rm{tclean}$ algorithm with Briggs weighting (robust=1.0), in which the beam size is $0.55{\times}0.47\arcsec$ (i.e., $9.5{\times}8.1$\,au). 
This weighting was selected here to achieve a beam size similar to the Band~6 image (naturally weighted, Fig.~\ref{fig:B6B7Image} bottom). 

The star should be at the centre of the image but has a positional uncertainty of ${\pm}0.05''$, due to the combination of the astrometric accuracy of \textit{Gaia} DR2 (${<}60\,\mu \rm{as}$) and \textit{ALMA} (which we assume to be ${<}10\%$ of the synthesised beam based on the \textit{ALMA} pipeline weblog ``timegaincal" plots; given our observation routine, choice of calibrators, and weather conditions, a reasonable target solution could be found). 
The peak detected at the centre of the image has a $\rm{S/N}{=}10.4$ and therefore could be from a source offset from its true position by $0.065''$, i.e., due to noise and systematic errors \citep[see Eq. 10.7][]{AlmaTH7}. 
As such the source detected at the centre of the image is consistent with the stellar position.

\begin{table}
    \centering
    \caption{Stellar position of q$^1$~Eri for the 6 observational epochs, calculated based on \textit{Gaia} DR2 data \citep{Gaia18}, for which DR2 positional errors are $<60\,\mu\rm{as}$. $a$: note that all Band~7 2016 data was collected on 12.07.16, so only one position of q$^1$~Eri is reported for this epoch.}
    \begin{tabular}{c|c|c|c|c}
         \hline
         \hline
         D.M.Y & RA & Dec\\
         \hline
         26.05.16 & 01:42:29.621 & -53:44:28.722 \\
         02.06.16 & 01:42:29.622 & -53:44:28.723 \\
         \hline
         12.07.16$^a$ & 01:42:29.624 & -53:44:28.735 \\ 
         \hline
         27.06.18 & 01:42:29.660 & -53:44:28.942 \\
         07.07.18 & 01:42:29.661 & -53:44:28.945 \\
         \hline
    \end{tabular}
    \label{tab:posGaia}
\end{table}

\begin{figure}
	\includegraphics[width=1.0\columnwidth]{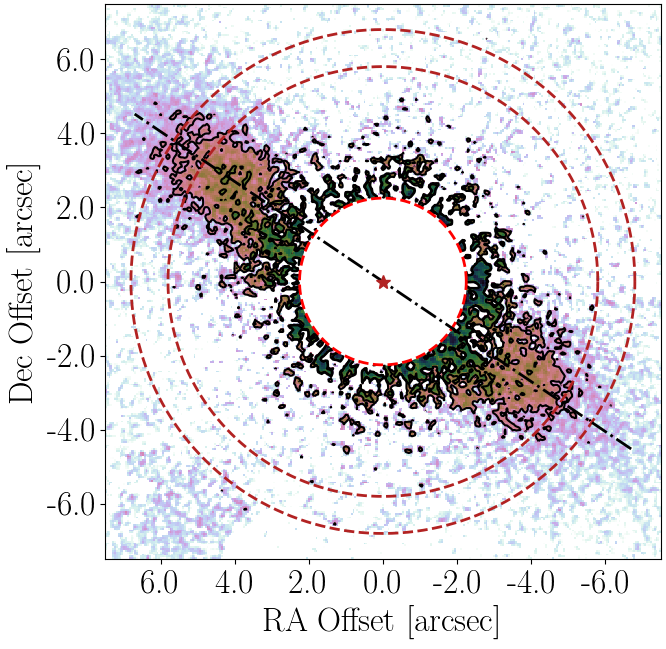}
    \caption{\textit{HST} image of q$^1$\,Eri, following F606W filtered observations (see details in Table~\ref{tab:HSTq1Obs}), showing contours for $+22$ and $+21$\,mag\,arcsec$^{-2}$ surface brightness. The inner red dashed ring indicates the scale of the coronagraph. The two outer dark-red dashed rings are at $5.7''$ and $6.7''$ from the stellar centre, showing that the emission in the NE is more extended than the SW along the major axis. To aid viewing the major axis, the black dash-dot line is shown with a position angle of $56^{\circ}$. In this figure, North is up, and East is left.}
    \label{fig:HST_alone}
\end{figure}

\subsection{Scattered Light HST Observations}
\label{sec:HSTObs}
q$^{1}$~Eri was observed by the coronagraph on-board the \textit{HST}'s Advanced Camera for Surveys (ACS) High Resolution Camera (HRC; $0.25''\,\rm{pixel}^{-1}$), as part of \textit{HST} program 10539 (PI: Karl Stapelfeldt), on 2nd September, 2006. 
Two observation sequences were made, as outlined in Table~\ref{tab:HSTq1Obs}, with the F606W filter (central wavelength: 5887\,\AA, FWHM: 1566\,\AA). 
Each sequence took one orbit in time, and the two sequences were made in consecutive orbits. 
Between the two sequences, the telescope was rolled about the line of sight axis by ${\sim}25^{\circ}$ to provide a mechanism to correct for systematic errors (such as instrumental artifacts) which can then be distinguished from real objects in the data. 
The ACS coronagraph reduces the wings of stellar diffraction patterns caused by the \textit{HST} aperture and obscurations. 
Any remaining halo then seen around the star is caused by optical surface errors (from scattering) that the coronagraph does not suppress. 

Since two images of q$^{1}$~Eri were collected at different roll angles, we used one as the reference point spread function (PSF) for the other. 
As the PSF is fixed on the detector, but the disc rotates with the roll angle, any sufficiently bright halo would appear as positive and negative signals when one image is subtracted from the other. 
While a face-on disc would ``self-subtract" with this method, q$^1$~Eri is sufficiently inclined for this method to be effective.
An algorithm was applied to iteratively solve for the sky and PSF images \citep[applied previously to detect debris discs by][]{Krist04}, producing a final sky image with residual halos. 
The final science image shown in Fig.~\ref{fig:HST_alone} is median filtered to remove cosmic rays, binned to 0.05$\arcsec$\,pix$^{-1}$ sampling, and shows the contour lines at the levels of +22 and +21\,mag\,arcsec$^{-2}$. 
This image clearly demonstrates the extent of q$^{1}$~Eri's debris disc \citep[and has been reduced identically as per][]{Stapelfeldt07} which we analyse further in $\S$\ref{sec:obsscatlight}. 

\begin{table}
    \centering
    \caption{\textit{HST} observing setup, taken over the two consecutive orbits. Subscripts 1 and 2 in `F606W$_{\rm{1}}$' and `F606W$_{\rm{2}}$' refer to the F606W filter being used for two different exposure times within a single observation sequence. The F502N automated acquisition exposures were taken with the star behind the coronagraph's occulting spot for 0.1s each, whilst the four F606W$_{\rm{2}}$ exposures were taken with the star centred behind the $1.8''$ diameter occulter.}
    \begin{tabular}{c|c|c|c|c|c}
         \hline
         \hline
         Orbit & F606W$_1$ [s] & F606W$_2$ [s] & Roll [$^{\circ}$] \\
         \hline
         1 & 540 (x4) = 2160 & 575 (x4) = 2300 & 0.0 \\
         2 & 540 (x4) = 2160 & 575 (x4) = 2300 & 24.9 \\
         \hline
    \end{tabular}
    \label{tab:HSTq1Obs}
\end{table}

\subsection{Flux Density Distribution}
\label{sec:SED}
\begin{figure}
	\includegraphics[width=\columnwidth]{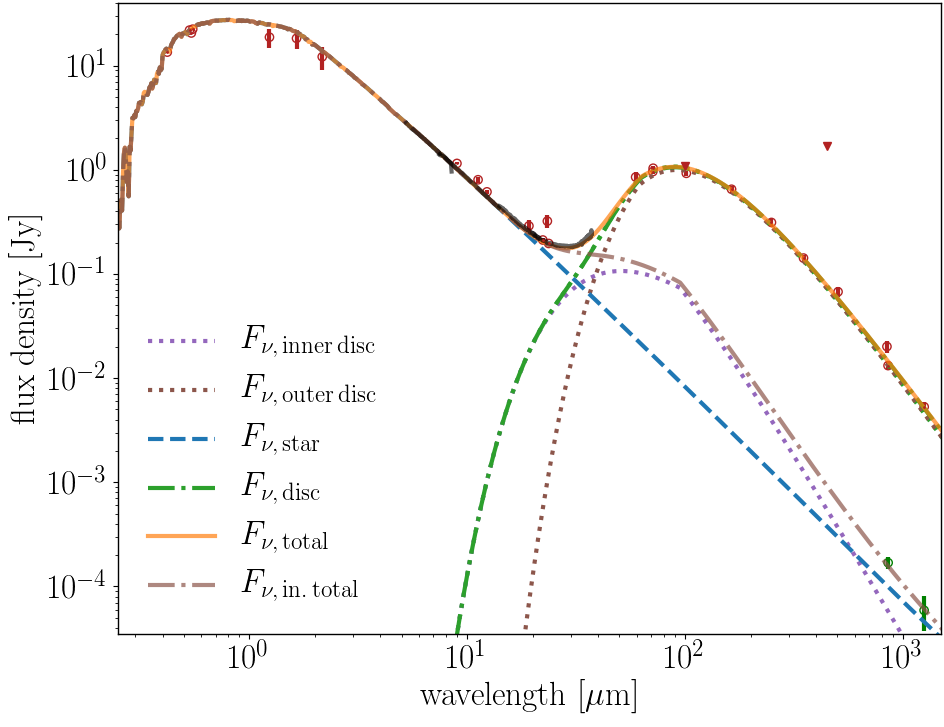}
    \caption{The flux density distribution of q$^{1}$~Eri, demonstrating the need to model this system with stellar emission, and both a cool (outer) and warm (inner) component. The warm and cool emission profiles (dotted lines) are modelled as modified blackbodies. The green dash-dot line shows the combination of both warm and cool components, and the the lilac dash-dot the combination of the stellar emission and warm component. Note that this data includes the derived flux from the \textit{ALMA} measurements outlined in $\S$\ref{sec:obsAnalysis}, and between ${\sim}5{-}36\,\mu$m the \textit{Spitzer IRS} spectra. }
    \label{fig:SED}
\end{figure}

Fig.~\ref{fig:SED} shows the flux density distribution of q$^{1}$~Eri, including fluxes derived from the \textit{ALMA} data in $\S$\ref{sec:obsAnalysis}. 
Values for the wavelengths and fluxes used to produce this are given in Table~\ref{tab:SEDflux}. 
The distribution was fitted using the methodology of \citet{Kennedy14} and \citet{Yelverton19}, finding a two-component temperature distribution (an inner warm component, and outer cool component) along with a stellar effective temperature of $T_{\rm{eff}}{=}6100{\pm}100\,\rm{K}$, and a stellar luminosity and radius (with 2\% calibration uncertainties) of $L{=}1.55\,L_{\odot}$, and $R{=}1.11\,R_{\odot}$ respectively. 
We note that \citet{Gaia18} estimated these stellar parameters (using the ``Priam" and ``FLAME" algorithms) as $6143\,\rm{K}$, $1.59\,L_{\odot}$, and $1.11\,R_{\odot}$ respectively, all broadly in agreement. 

The two-component model finds an inner warm belt and an outer cool belt, respectively with blackbody temperatures of $T_{\rm{bb,inner}}{=}101{\pm}4\,\rm{K}$ and $T_{\rm{bb,outer}}{=}41{\pm}1\,\rm{K}$ (corresponding to uncorrected blackbody radii of $9.5{\pm}0.7$\,au and $57{\pm}3$\,au). 
Respectively these components have fractional luminosities of $6.3{\pm}0.4{\times}10^{-5}$ and $2.44{\pm}0.08{\times}10^{-4}$. 
Although due to non-photospheric emission and the possibility of additional unresolved circumstellar material such as hot dust being located near the star \citep[and thus a strict blackbody extrapolation may not be an accurate representation of the stellar spectrum, see][]{White20}, we can nevertheless use these models to estimate the $856\,\mu$m and 1.25\,mm emission of the star, and find these to be $99\,\mu$Jy and ${\sim}46\,\mu$Jy, both with 2\% uncertainty respectively. 

\begin{table}
    \centering
    \caption{Values for the fluxes used to produce the flux density distribution in Fig.~\ref{fig:SED}. a= \citet{2006yCat.2168....0M}, b= \citet{2015A&A...580A..23P}, c= \citet{2000A&A...355L..27H}, d= \citet{Gaia18}, e= \citet{ESA97}, f= \citet{2006AJ....131.1163S}, g= \citet{2003tmc}, h= \citet{2010AJ....140.1868W}, i= \citet{2010A&A...514A...1I}, j= \citet{1988iras6A}, k= \citet{2010ApJS..191..212S}, l= \citet{2018MNRAS.475.3046S}, m= \citet{2017MNRAS.470.3606H}, n= \citet{2015ApJ...813..138R}, o= this work. All wavelengths are provided to 3 significant figures. The calculation of the \textit{ALMA} values in the lower part of this table are derived in $\S$\ref{sec:obsAnalysis}.}
    \begin{tabular}{c|c|c|c}
         \hline
         \hline
         Source & Wavelength & Flux /  & Units \\
         & [$\mu\rm{m}$] & Magnitude & \\
         \hline
         U-B $^{\rm{a}}$ & - & $0.00{\pm}0.12$ &mag \\
         C$_1$ $^{\rm{b}}$ & - & $0.336{\pm}0.012$ &mag \\
         B$_{\rm{T}}$ $^{\rm{c}}$ & 0.420 & $6.164{\pm}0.014$ &mag \\
         M$_1$ $^{\rm{b}}$ & - & $0.168{\pm}0.008$ &mag \\
         B-V $^{\rm{a}}$ & - & $0.530{\pm}0.026$ &mag \\
         B-Y $^{\rm{b}}$ & - & $0.354{\pm}0.008$ &mag \\
         B$_{\rm{P}}$ $^{\rm{d}}$ & 0.513 & $5.6770{\pm}0.0017$ &mag \\
         V$_{\rm{T}}$ $^{\rm{c}}$ & 0.532 & $5.581{\pm}0.009$ &mag \\
         H$_{\rm{P}}$ $^{\rm{e}}$ & 0.542 & $5.638{\pm}0.006$ &mag \\
         V $^{\rm{a}}$ & 0.550 & $5.540{\pm}0.019$ &mag \\
         G $^{\rm{d}}$ & 0.642 & $5.3550{\pm}0.0013$& mag \\
         R$_{\rm{P}}$ $^{\rm{d}}$ & 0.780 & $4.9660{\pm}0.0022$ &mag \\
         I $^{\rm{f}}$ & 0.791 & $8.757{\pm}0.020$ &mag \\
         J $^{\rm{g}}$ & 1.24 & $4.79{\pm}0.23$ &mag \\
         H $^{\rm{g}}$ & 1.65 & $4.40{\pm}0.23$ &mag \\
         K$_{\rm{S}}$ $^{\rm{g}}$ & 2.16 & $4.34{\pm}0.28$ &mag \\
         \textit{WISE} W1 $^{\rm{h}}$ & 3.38 & $4.17{\pm}0.38$ &mag \\
         \textit{WISE} W2 $^{\rm{h}}$ & 4.63 & $3.91{\pm}0.22$ &mag \\
         \textit{AKARI} IRC $^{\rm{i}}$ & 8.98 & $1.151{\pm}0.019$ &Jy \\
         \textit{IRAS} $^{\rm{j}}$ & 11.2 & $0.82{\pm}0.06$ &Jy \\
         \textit{WISE} W3 $^{\rm{h}}$ & 12.3 & $4.22{\pm}0.014$ &mag \\
         \textit{AKARI} IRC $^{\rm{i}}$ & 19.2 & $0.312{\pm}0.039$ &Jy \\
         \textit{WISE} W4 $^{\rm{h}}$ & 22.3 & $3.954{\pm}0.021$ &mag \\
         \textit{IRAS} $^{\rm{j}}$ & 23.3 & $0.34{\pm}0.04$ &Jy \\
         \textit{MIPS} $^{\rm{k}}$ & 23.7 & $196.20{\pm}0.08$ &mJy \\
         \textit{IRAS} $^{\rm{j}}$ & 59.4 & $0.85{\pm}0.11$ &Jy \\
         \textit{PACS} $^{\rm{l}}$ & 71.2 & $0.961{\pm}0.010$ &Jy \\
         \textit{MIPS} $^{\rm{k}}$ & 71.4 & $1035{\pm}6$ &mJy \\
         \textit{IRAS} $^{\rm{j}}$ & 100 & $<1.08$ &Jy \\
         \textit{PACS} $^{\rm{l}}$ & 101 & $0.925{\pm}0.008$ &Jy \\
         \textit{PACS} $^{\rm{l}}$ & 164 & $0.651{\pm}0.018$ &Jy \\
         \textit{SPIRE} $^{\rm{l}}$ & 249 & $0.312{\pm}0.026$ &Jy \\
         \textit{SPIRE} $^{\rm{l}}$ & 350 & $0.1421{\pm}0.0038$ &Jy \\
         \textit{JCMT} $^{\rm{m}}$ & 447 & $<1701$ &mJy \\
         \textit{SPIRE} $^{\rm{l}}$ & 504 & $0.0674{\pm}0.0035$& Jy \\
         \textit{JCMT} $^{\rm{m}}$ & 845 & $20.1{\pm}2.7$ &mJy \\
         ATCA $^{\rm{n}}$ & 6760 & $0.093{\pm}0.017$ &mJy \\
         \hline
         \textit{ALMA} B7 $^{\rm{o}}$ & 856 & $13.2 {\pm} 1.3$ &mJy \\
         \textit{ALMA} B7 (inner) $^{\rm{o}}$ & 856 & $169 {\pm} 22$ & $\mu$Jy \\
         \textit{ALMA} B6 $^{\rm{o}}$ & 1250 & $5.3 {\pm} 0.6$ &mJy \\
         \textit{ALMA} B6 (inner) $^{\rm{o}}$ & 1250 & $59 {\pm} 22$ & $\mu$Jy \\
         \hline
    \end{tabular}
    \label{tab:SEDflux}
\end{table}

We note here that since the emission from the region consistent with the stellar location is resolved separately to the emission due to the complete q$^1$~Eri system, data points can be plotted for this inner emission (see green data points in lower-right of Fig.~\ref{fig:SED}). 
Using modified blackbodies, these data points allowed us to constrain the inner component more tightly than would have been possible if the images of q$^1$~Eri were unresolved. 
Our modified blackbodies are identical to regular blackbodies until a critical wavelength, $\lambda_0$, after which the flux density is multiplied by $(\lambda/\lambda_0) ^{-\beta_0}$. 
The temperature (and thus blackbody radius) of this inner component is relatively well constrained (i.e., by near and mid-infrared emission near this component's peak). 
However, the parameters defining the blackbody modification are not well constrained (i.e., due to uncertainty on the stellar photosphere and given that the two Band~6 and Band~7 data points these model have only low signal-to-noise excesses) and thus there are a broad range of values for which $\lambda_0$ and $\beta_0$ are consistent with our data. 
For this reason, whilst this two-component model gives a good match to the complete flux distribution of q$^1$~Eri, and the Band~6 and 7 data points in the inner regions of the system, there still remains large uncertainty on what might be expected for the faint sub-mm emission of the inner region at these wavelengths. 
We discuss the implications of this further in $\S$\ref{sec:radialProfAn} and $\S$\ref{sec:discussionWarm}. 

We note that there are two ${\sim}$850\,$\mu$m flux measurements, our ALMA data (13.2${\pm}$1.3mJy) and JCMT data (20.1${\pm}$2.7mJy). 
These measurements are formally consistent with each other to within $2.3\,\sigma$, but we prefer the ALMA measurement since it is consistent with the Band~6 measurement for a typical debris disc millimeter spectral index of 2.5 \citep{Ricci15, MacGregor16}. 
This consistency also suggests that the ALMA observations in the two bands with differing maximum recoverable scales does not resolve out significant levels of emission.

\section{Observational Analysis}
\label{sec:obsAnalysis}
In this section we present an initial analysis of the observations outlined in $\S$\ref{sec:ALMAObsFull} to assess the disc's morphology and any asymmetries, motivating the more detailed modelling in $\S$\ref{sec:modelling}.

\subsection{Continuum Analysis}
\begin{figure}
	\includegraphics[width=1.0\columnwidth]{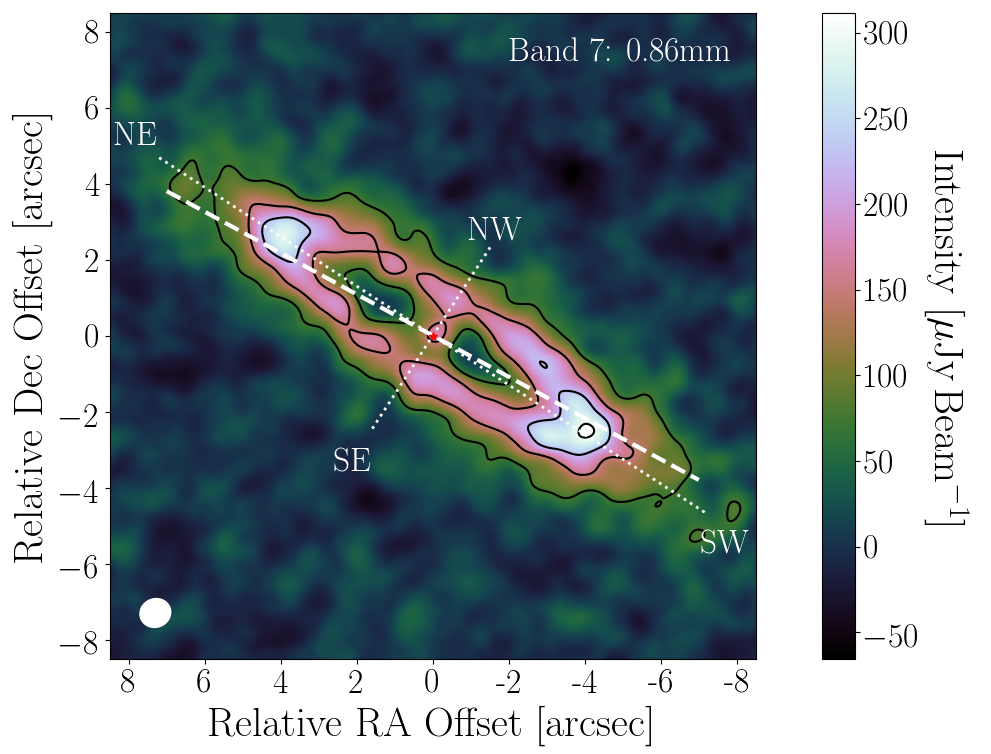}
    \caption{The tapered \textit{ALMA} Band~7 image (natural weighted), demonstrating the new beam size of ${\sim}0.82{\times}0.75''$ in the lower left in white, the intensity of the disc emission with contours of +5, +10, +15 and +20$\,\sigma$ significance, the directions of the major axis (NE-SW) and minor axis (NW-SE) with white dotted lines, the stellar location in the image centre as a red star, and the line dividing the disc in half by integrated flux is the white dashed line (see later $\S$\ref{sec:radialProfAn}). }
    \label{fig:ALMAB7Tap}
\end{figure}

Prior to conducting any further image analysis we raised the S/N per beam (by ${\sim}50\%$) of the Band~7 data by applying a uv-taper of $0.5''$, effectively increasing the synthesised beam to ${\sim}0.82\times0.75''$, shown in Fig.~\ref{fig:ALMAB7Tap}. 
This raises the S/N in comparison to the Briggs weighted image shown in the top of Fig.~\ref{fig:B6B7Image}. 
Throughout this work the directions and disc positions referred to as the North-East (NE), South-East (SE), North-West (NW), and South-West (SW) are shown in this diagram. 
We also show the major (NE-SW) and minor (NW-SE) axes shown as white dotted lines (based on the position angle of the disc, calculated in the following paragraph). 

We first measure the disc position angle, $\rm{PA}$ (measured anti-clockwise from North) and inclination, $i$ (for which $90^{\circ}$ would be edge-on), required to deproject the data. 
We found the position angle by plotting the total flux within $10^{\circ}$ wedges from the stellar centre as a function of angle (iterating this procedure from $0$ to $180^{\circ}$ in $1^{\circ}$ increments), which finds a value of $\rm{PA}{=}57.0{\pm}1.0^{\circ}$, consistent with previous analyses which found the position angle as ${\sim}56^{\circ}$ and $54{\pm}5^{\circ}$ respectively \citep[see][]{Stapelfeldt07, Liseau10}. 
We define the direction of the position angle as the disc major axis, the minor axis as the line perpendicular to this, both through the stellar position.  

\begin{figure}
	\includegraphics[width=1.0\columnwidth]{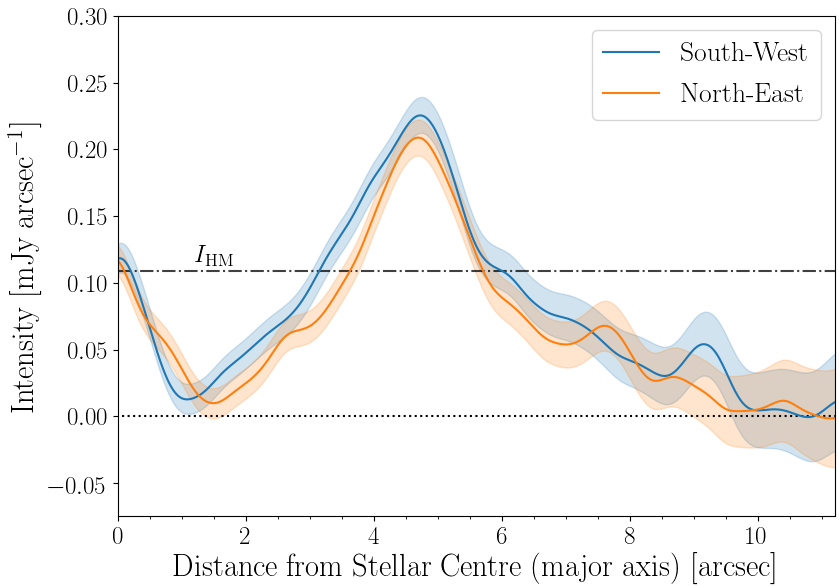}
	\includegraphics[width=1.0\columnwidth]{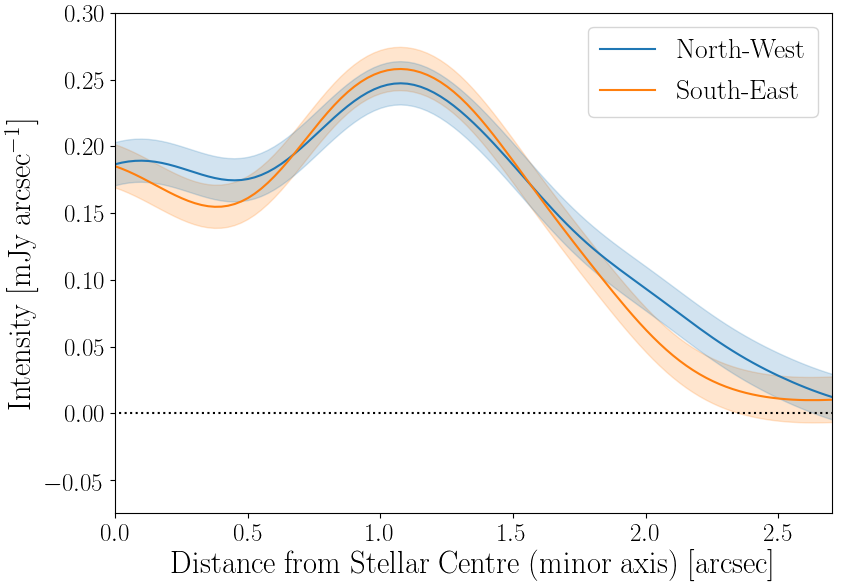}
    \caption{Integrated radial profiles for the Band 7 image in Fig.~\ref{fig:ALMAB7Tap}, taken along the major axis (top) and minor axis (bottom), with $I_{\rm{HM}}$, the half-maximum intensity, shown on the top plot. The shaded regions show the $1\,\sigma$ error. Our procedure for producing these is outlined in the first paragraph of $\S$\ref{sec:radialProfAn}.}
    \label{fig:ALMAB7RadProfs}
\end{figure}

\subsubsection{Continuum Brightness Analysis}
\label{sec:radialProfAn}
Fig.~\ref{fig:ALMAB7RadProfs} shows the radial profiles (brightness as a function of distance from the star) in the major and minor axes (upper and lower plots respectively) for the Band~7 image. 
These are measured by summing the flux in pixels in columns perpendicular to the plotted axis. 
For the major and minor profiles, these are found in columns either ${\sim}0.3\arcsec$ or ${\sim}0.5\arcsec$ (respectively) above or below the axis being plotted, to minimise signal from off-axis emission. 
The peak brightnesses and associated radii found from these profiles are given in Table~\ref{tab:radProfVals}, along with the average peak to peak radial distances. 
This shows that the disc emission brightness peaks at an average projected distance of $81.6{\pm}0.5$\,au from the star, that the peak brightnesses in the major axis (NE and SW) are consistent, and that the peak brightnesses in the minor axis (NW and SE) are also consistent (i.e., based on these peaks, the disc geometry is consistent with a circle centred on the star). 
To estimate the inclination, $i$, we used the peak to peak radial averages ($r_{\rm{maj, peak-peak}}{=}4.71{\pm}0.03''$ and $r_{\rm{min, peak-peak}}{=}1.08{\pm}0.07''$), since for rings with low eccentricity, $i {\approx} \arccos(r_{\rm{min, peak-peak}} / r_{\rm{maj, peak-peak}})$, giving $i{=}76.7{\pm}1.0^{\circ}$, also consistent with previous analyses which found this inclination to be $76^{\circ}$ and ${>}63^{\circ}$ respectively \citep[see][]{Stapelfeldt07, Liseau10}. 
To attain projected radii (in au) from our radial profiles in Fig.~\ref{fig:ALMAB7RadProfs} (as noted in Table~\ref{tab:radProfVals}), we assume $i{=}77^{\circ}$.

\begin{table}
    \centering
    \caption{Integrated radial profile derived radii, inner and outer edges, widths, brightnesses and radial offsets from Fig.~\ref{fig:ALMAB7RadProfs}. Combined errors are found from the quadrature sum of the individual measurement values for the averages. Widths in the major axis measurements are defined as the difference in distance between the inner and outer edges. Averages found around the ring are detailed in the lower panel of the table. Projected distances in au have assumed an inclination of $i{=}77^{\circ}$ and the \textit{Gaia} distance to q$^1$~Eri of 17.34\,pc.}
    \begin{tabular}{c|c|c}
         \hline
         \hline
         Major Axis & Direction & Value \\
         \hline
         Peak Radius & NE & $4.70{\pm}0.04''$ | $81.5{\pm}0.7$\,au\\
         Peak Radius & SW & $4.71{\pm}0.04''$ | $81.7{\pm}0.7$\,au\\
         Peak Brightness & NE & $0.225{\pm}0.013$\,mJy\,arcsec$^{-1}$ \\
         Peak Brightness & SW & $0.209{\pm}0.013$\,mJy\,arcsec$^{-1}$ \\
         Inner Edge & NE & $3.62{\pm}0.14''$ | $62.8{\pm}2.4$\,au\\
         Outer Edge & NE & $5.71{\pm}0.18''$ | $99.0{\pm}3.1$\,au\\
         Inner Edge & SW & $3.14{\pm}0.14''$ | $54.4{\pm}2.4$\,au\\
         Outer Edge & SW & $6.02{\pm}0.35''$ | $104{\pm}6$\,au\\
         Width$_{\rm{NE}}$ & NE & $2.09{\pm}0.22''$ | $36.2{\pm}3.8$\,au\\
         Width$_{\rm{SW}}$ & SW & $2.88{\pm}0.38''$ | $50{\pm}7$\,au\\         
         \hline
         Minor Axis & Direction & Value \\
         \hline
         Peak Radius & NW & $1.08{\pm}0.10''$ | $77.4{\pm}7.2$\,au\\
         Peak Radius & SE & $1.08{\pm}0.10''$ | $77.4{\pm}7.2$\,au\\
         Peak Brightness & NW & $0.247{\pm}0.016$\,mJy\,arcsec$^{-1}$ \\
         Peak Brightness & SE & $0.258{\pm}0.016$\,mJy\,arcsec$^{-1}$ \\
         \hline
         Averages & Axis & Value \\
         \hline
         Peak-Peak Radius & Major & $4.71{\pm}0.03''$ | $81.6{\pm}0.5$\,au\\
         Peak-Peak Radius & Minor & $1.16{\pm}0.07''$ | $77.4{\pm}5.0$\,au\\
         Inner Edge & Major & $3.38{\pm}0.10''$ | $58.6{\pm}1.7$\,au\\
         Outer Edge & Major & $5.87{\pm}0.20''$ | $101.8{\pm}3.5$\,au\\
         Width ($W_{\rm{HM}}$) & Major & $2.49{\pm}0.22''$ | $43{\pm}4$\,au\\
         Peak Brightness & Major & $0.217{\pm}0.009$\,mJy\,arcsec$^{-1}$ \\
         Peak Brightness & Minor & $0.253{\pm}0.011$\,mJy\,arcsec$^{-1}$ \\
         \hline
         Radial Offset & Axis & Value \\
         \hline
         $R_{\rm{inner}}$ & Major & $0.071{\pm}0.029$ \\
         $R_{\rm{peak}}$ & Major & $-0.001{\pm}0.006$ \\
         $R_{\rm{outer}}$ & Major & $-0.026{\pm}0.033$ \\
         \hline
    \end{tabular}
    \label{tab:radProfVals}
\end{table}

We firstly define some of the features of the major axis profile (top) in Fig.~\ref{fig:ALMAB7RadProfs}, which is more clearly resolved and has a higher signal to noise than the minor axis profile. 
This profile demonstrates that there is no significant difference between the peak brightness in the NE and SW directions, nor is there a difference in the radius at which this peak is measured. 
Calculating the average of the two NE and SW emission peak intensities and halving this, we also measured radii consistent with this half-maximum intensity, which we use as a measure of the disc width, and find this to be $W_{\rm{HM}}{=}43{\pm}4$\,au (see values in Table~\ref{tab:radProfVals}, under ``Averages"). 
This width measurement is affected less by noisy emission towards the outer regions of the image, and we therefore use the radial location of the half-peak emission radii to define the inner and outer edge locations in both directions, and the width of the disc ansa in the NE and SW between these radii, respectively. 
We note however that if the average radius in both NE and SW directions at which emission exceeds $3\,\sigma$ is used to define its extent, then the disc emission can be seen to extend over a broader $34{-}134$\,au. 
Since the image is signal-to-noise limited, this is a lower limit, and the disc extends over at least this range. 

Whereas the radial profiles beyond the emission peak (i.e., on the outer side) in the NE and SW directions are consistent within their error bars for all measured radii, this is not the case on the inner side, where there is a significant radial width over which the South-West is brighter than the North-East. 
We quantify the extent a given radial diagnostic $r$ (i.e., the projected distance from the star) differs between the NE and SW (in $r_{\rm{NE}}$ and $r_{\rm{SW}}$) using what we call the radial offset, $R {=} (r_{\rm{NE}}-r_{\rm{SW}}) / (r_{\rm{NE}}+r_{\rm{SW}})$, for which a positive value would indicate that the emission measured is offset from the stellar emission (at the coordinate centre) in the NE direction. 
The offset of the inner edge $R_{\rm{inner}} {=} 0.071{\pm}0.029$ is $2.4\,\sigma$ significant, but those of the location of the peak emission and of the outer edge are insignificant of $R_{\rm{peak}} {=} -0.001{\pm}0.006$ and $R_{\rm{outer}} {=} -0.026{\pm}0.033$. 
Although the significance of $R_{\rm{inner}}$ does not exceed $3\,\sigma$, the SW emission exceeds the NE emission for all inner edge radii ${\gtrsim}1.2\arcsec$, and therefore there are a broad range of radii on the inner edge that such an offset measurement would yield a value larger than 0 (i.e., $R_{\rm{inner}}$ can be found at a similar level over a radial extent that covers several beam widths). 

We considered the minor axis emission similarly to the major axis, as plotted in the lower profile of Fig.~\ref{fig:ALMAB7RadProfs}, and tabulated data in Table~\ref{tab:radProfVals}. 
We find that the peak brightnesses do not differ significantly between the NW and SE and that the peak emission radii are strongly consistent, however in the minor axis there is no evidence of a radial offset at any radii.

\begin{figure}
    \centering
    \includegraphics[width=1.0\columnwidth]{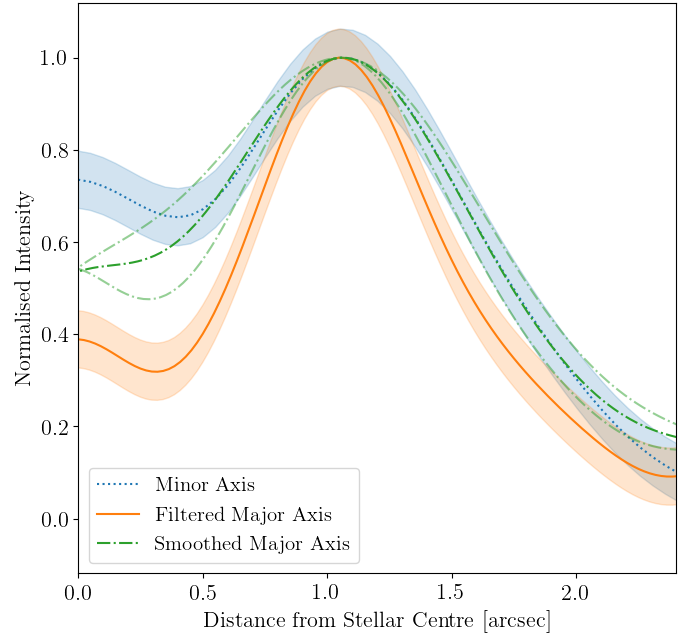}
    \caption{Normalised emission intensity as a function of the distance from the stellar centre for the 'filtered' major axis profile (average of the SW and NE profiles, projected onto the minor axis following deconvolution and convolution to account for the change in beam extent, in solid, amber), the filtered major axis smoothed with a further Gaussian convolution (in dashed, green), and the average minor axis intensity profile (in dotted, blue).}
    \label{fig:scaleHeight}
\end{figure}

Most debris discs with resolved scale heights are viewed edge-on. 
When considering the narrower and less inclined debris disc of HD\,181327, \citet{Marino16} showed that inclined discs ($i{\sim}70^{\circ}$) observed with \textit{ALMA} have measurably different azimuthal brightness profiles for low and moderate scale heights. 
Given that we have determined q$^1$~Eri's debris disc to be more inclined than HD\,181327, we therefore investigate whether or not the dimensionless vertical aspect ratio $h=H/r$ is resolved. 
Herein we refer to the vertical aspect ratio as the scale height, and define $H$ as the height of dust above the midplane at a radius $r$. 

We first note that the major axis radial profile would be largely unaffected by the vertical emission distribution, which would not be the case in the minor axis of an inclined disc. 
We then use the major axis radial profile to determine what the minor axis radial profile should look like if the disc was flat (i.e., $h{=}0$), accounting for the inclination projection and the smoothing produced by the beam size parallel with the minor axis (i.e., by re-scaling the major axis profile to compare with the minor axis). 
In Fig.~\ref{fig:scaleHeight} we plot this comparison, referring to the major axis profile viewed in the minor axis as the filtered major axis (assuming this to be thin), alongside the averaged minor axis radial profile. 
We find that the observed minor axis profile is broader than the filtered major axis profile (if assumed to be thin) which would be the case if the disc instead has a vertical scale height. 

By convolving the filtered major axis profile with a Gaussian ($\sigma_{\rm{h}}{=}0.8\arcsec$), we broadly reproduce the minor axis profile, shown on Fig.~\ref{fig:scaleHeight} as the smoothed major axis profile. 
We also plot above and below the smoothed major axis profile, the result of convolving the filtered major axis instead with a Gaussian, each with $\sigma_{\rm{h}}{=}0.6\arcsec$ (lower) and $\sigma_{\rm{h}}{=}1.0\arcsec$ (upper). 
Since these also broadly fit the width of the $1\,\sigma$ error region of the minor axis, this allows us to constrain the level of smoothing required, which we find to be consistent with that introduced by emission with a constant scale height $h{=}0.04{\pm}0.01$. 
Therefore, whilst this may only be weakly constrained (and this may be biased by other sub-structure in the disc), such an analysis tentatively suggests that the vertical scale height of q$^1$~Eri has been resolved in Band~7, and that this is at the level of a few percent of the disc radius. 

We finally note that the peak emission coincident with the star, if assumed to be a point source, has a Band~7 flux of $F_{\rm{star,B7}}{=}169 {\pm}22\, \mu \rm{Jy}$ and a Band~6 flux of $F_{\rm{star,B6}}{=}59 {\pm}15\, \mu \rm{Jy}$, with the error estimated by summing in quadrature the image rms with an assumed 10\% flux calibration uncertainty. 
For the Band~7 emission, this is higher than the ${\sim} 99$\,$\mu$Jy expected from the star by ${\sim}3.2\,\sigma$, although the ${\sim} 46$\,$\mu$Jy Band~6 emission is consistent with the stellar emission (see Table~\ref{tab:fluxVals} and \S \ref{sec:SED}). 
This Band~7 excess may therefore comprise of emission from the stellar photosphere and an extra unresolved warm emission component, from a possible inner belt, which we discuss further in \S \ref{sec:discussionWarm}. 

In summary, by analysing the brightness distribution of the Band~7 image, we have demonstrated that this disc is inclined, broad in extent, has a significant radial offset towards the SW (in the major axis) on the disc inner edge, is symmetric in the minor axis, and may contain detectable sub-mm emission from an inner warm component near to the star. 
Such a major axis offset could be the result of a larger-scale asymmetric distribution in the disc, such as the presence of a clump on the SW inner edge, due to the distribution being eccentric (with a pericentre direction between the NW and SW), a combination of these, or simply due to noise. 
We explore these hypotheses further in $\S$\ref{sec:FPAObsAn}, \ref{sec:modFitResComp} and \ref{sec:discOriginAsymms}. 

\begin{table}
    \centering
    \caption{Table of flux values calculated within the defined regions from Fig.~\ref{fig:fluxNESW} and Fig.~\ref{fig:fluxNESW_b6}, including the flux ratios comparing the two disc halves. Note that a $10\%$ flux calibration error has been added in quadrature to the determined $F_{\rm{tot}}$ values.}
    \begin{tabular}{c|c|c}
         \hline
         \hline
         Flux measurements & Band~7 & Band~6 \\
         & [$\rm{mJy}$] & [$\rm{mJy}$] \\
         \hline
         $F_{\rm{star}}$ & $0.169{\pm}0.022$  & $0.059{\pm}0.015$  \\
         $F_{\rm{tot}}$ & $13.2{\pm}1.3$  & $5.3{\pm}0.6$  \\
         $F_{\rm{SW,\,82au}}$ & $1.78{\pm}0.06$  & $0.74{\pm}0.06$  \\
         $F_{\rm{SW,\,200au}}$ & $7.23{\pm}0.09$  & $2.59{\pm}0.09$  \\
         $F_{\rm{NE,\,82au}}$ & $1.48{\pm}0.06$  & $0.67{\pm}0.06$  \\
         $F_{\rm{NE,\,200au}}$ & $5.90{\pm}0.09$  & $2.59{\pm}0.09$  \\
         $\delta F$ & $1.33{\pm}0.13$  & $0.00{\pm}0.13$  \\
         \hline
         Flux ratios & Band~7 & Band~6 \\
         \hline
         $f_{\rm{82au}}$ & $1.21{\pm}0.06$ & $1.10{\pm}0.13$ \\
         $f_{\rm{200au}}$ & $1.23{\pm}0.04$ & $1.00{\pm}0.07$ \\
         \hline
    \end{tabular}
    \label{tab:fluxVals}
\end{table}

\subsubsection{Flux Profile Analysis}
\label{sec:FPAObsAn}
\begin{figure}
    \includegraphics[width=1.0\columnwidth]{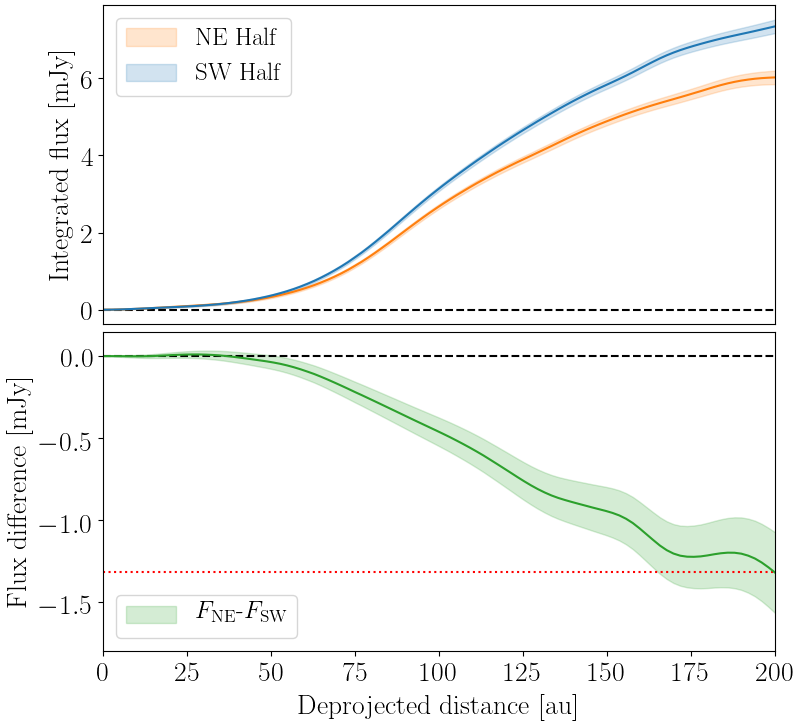}
    \caption{Band~7 flux profiles for the inner $200$\,au of q$^1$~Eri \textit{ALMA} data, found by integrating the total flux within deprojected radial bins cumulatively. Top: Flux profile for the full image, with shaded regions showing the ${\pm}1\,\sigma$ errors. Bottom: Difference profile for the full image, for flux in the NE half minus flux in the SW half. The shaded region is the ${\pm}1\,\sigma$ quadrature sum of the error from the two halves, and the red-dotted line shows the flux difference at 200\,au.}
    \label{fig:fluxNESW}
\end{figure}

\begin{figure}
    \includegraphics[width=1.0\columnwidth]{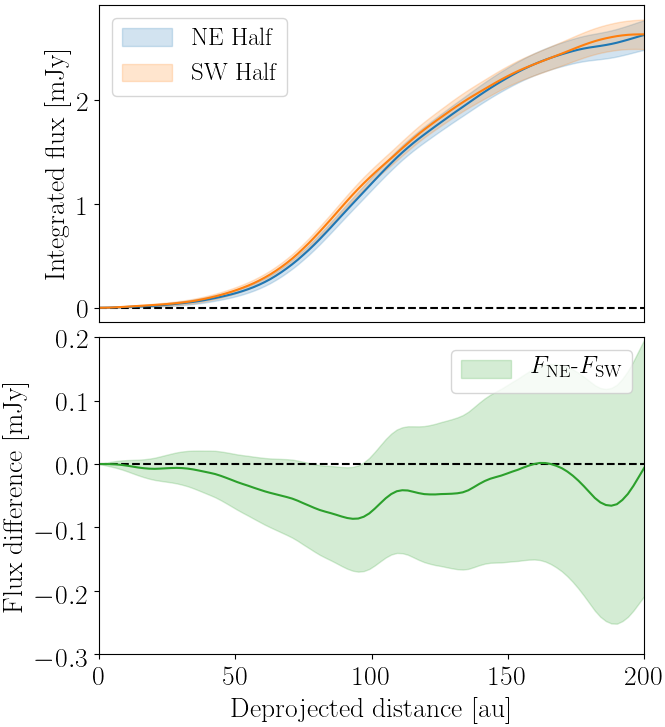}
    \caption{As per Fig.~\ref{fig:fluxNESW} but for the Band~6 data.}
    \label{fig:fluxNESW_b6}
\end{figure}

We produced flux profiles for the image by halving the Band~7 image (Fig.~\ref{fig:ALMAB7Tap}) along the minor axis, and summing the total flux within projected radial bins either side of the minor axis. 
The results are shown for the Band~6 and 7 images in Figs.~\ref{fig:fluxNESW} and~\ref{fig:fluxNESW_b6}. 
Table~\ref{tab:fluxVals} shows the values measured from these plots of the total flux, $F_{\rm{tot}}$, within a projected distance of $200$\,au from the star. 
These are used to produce the flux distribution in Fig.~\ref{fig:SED}, and can be used to calculate the spectral index $\alpha_{\rm{mm}}{=}2.34{\pm}0.29$. 
Given that the previous brightness profile analysis concluded that an asymmetry may exist on the inner edge of the disc, we sought to quantify this further with measurements of the integrated flux. 
By ignoring emission internal to 30\,au (i.e., from the stellar photosphere and possible inner warm component), we measured the total flux on either side of the disc internal to the brightness maxima radii (i.e., internal to ${\sim}82$\,au), noted as $F_{\rm{SW,\,82\,au}}$ and $F_{\rm{NE,\,82\,au}}$ respectively, and internal to 200\,au, noted as $F_{\rm{SW,\,200\,au}}$ and $F_{\rm{NE,\,200\,au}}$ respectively, and also the net difference between these at 200\,au, $\delta F = F_{\rm{SW,\,200\,au}} - F_{\rm{NE,\,200\,au}}$. 
To quantify the flux enhancement on either side of the disc, we also calculated the flux ratios between the NE and SW from these same measurements, as $f_{\rm{82\,au}}{=}F_{\rm{SW,\,82\,au}}/F_{\rm{NE,\,82\,au}}$ and $f_{\rm{200\,au}}{=}F_{\rm{SW,\,200\,au}}/F_{\rm{NE,\,200\,au}}$ respectively. 
We note here that whilst the absolute values of fluxes should all include flux calibration uncertainties of $10\%$, this uncertainty does not feature in the calculations for flux ratios, since this calibration uncertainty affects all flux in this map equally (i.e., to assess asymmetries this does not need to be included, but for individual flux measurements this does).

Considering first the Band~7 data, it can be seen in Fig.~\ref{fig:fluxNESW} that although the shapes of the integrated flux in the NE and SW are similar, the values diverge. 
This is more easily seen in the lower plot which quantifies the difference between these. This shows that from ${\sim}50$\,au (i.e., on the inner edge) these are significantly different, reaching a maximum difference at ${\sim}200$\,au. 
Although the peak flux difference value is found at this longest radius, this is consistent with no flux difference beyond ${\sim}160$\,au, i.e., on the outer extent of the disc (see $\S$\ref{sec:radialProfAn}). 
We quantify this difference in Table~\ref{tab:fluxVals} as $\delta F {=} 1.33{\pm}0.18$\,mJy, and the flux ratios along the inner edge and over the full disc extent as $f_{\rm{82au}}{=}1.21{\pm}0.06$ and $f_{\rm{200au}}{=}1.23{\pm}0.04$ respectively, all of which show strong evidence that the Band~7 disc flux is asymmetric. 
The Band~6 flux profile in Fig.~\ref{fig:fluxNESW_b6} shows no significant flux asymmetry. 
Although the lower plot demonstrates that the SW may contain more flux, this is not significant, as quantified by the Band~6 $\delta F$ value (consistent with 0) or either the $f_{\rm{82\,au}}$ and $f_{\rm{200\,au}}$ values (both consistent with 1). 
Despite these Band~6 flux ratios and flux difference not showing evidence for an asymmetry, the lower S/N of this image means that an asymmetry could still be present at broadly the same ${\sim}20$ percent level seen in Band~7. 
Deeper observations of q$^1$~Eri in Band~6 would be required to determine this.

\subsubsection{Interpreting the Flux and Brightness Profiles}
\label{sec:obsInts}
\begin{figure}
    \includegraphics[width=1.0\columnwidth]{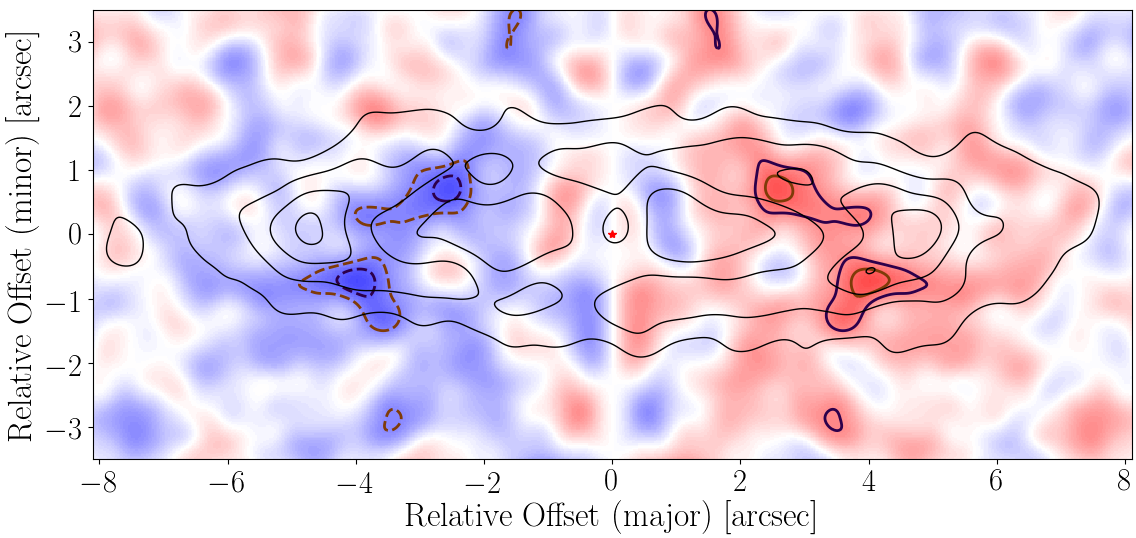}
    \caption{A minor axis mirror-subtracted Band~7 image of q$^1$~Eri where the flux in each pixel has had subtracted that of the corresponding pixel on the opposite side of the minor axis at 0'' offset (vertical in image). This also shows the continuum emission contour lines for +6, +11 and +16$\,\sigma$, and the significance of the emission after subtraction in the blue and red contours, respectively ${\pm}3$ and ${\pm}4\,\sigma'$ (where $\sigma'{=}\sqrt{2}\sigma_{\rm{B7\,image}}$). Note that the left and right side of this image represent the North-Eastern and South-Western halves of the disc respectively. }
    \label{fig:mirrorSub}
\end{figure}

If the disc was overall symmetrical, but with a clump in the SW ansa, then this would lead to a brightness asymmetry between the two ansae. 
To explore this, Fig.~\ref{fig:mirrorSub} shows a mirror-subtracted image in which the flux in each pixel has had subtracted that of the corresponding pixel on the opposite side of the minor axis (i.e., relative to $0\arcsec$ offset in the major axis). 
This plot shows the thin black contour lines of the original Band~7 image, and coloured contours where the mirror subtraction has resulted in significant residual emission (i.e., ${>}3\,\sigma$). 
This also demonstrates that the asymmetry in the SW ansa is interior to the peak brightness radius and may extend over a broad azimuthal range of the disc. 
We note that the two $4\,\sigma$ residual contours have sizes of ${\approx}{0.7''}$ respectively at projected orbital radii ${\sim}{60}$\,au.
No further asymmetries are evident radially beyond the emission peak which are coincident with the disc, consistent with our analysis of the brightness. 
Thus, one scenario that we will investigate in $\S$\ref{sec:modelling} with model \textbf{CL} is that the sub-mm emission arises from a broad, symmetric ring of parent planetesimals with an extended clump of emission on the inner edge of the SW ansa (equivalently, this could be interpreted as the SW ansa being radially broader than the NE). 

Another interpretation of the asymmetries that we will explore with model \textbf{ECC} is that these are not caused by a clump, but due to the disc being eccentric. 
If this was the case, the disc flux would be symmetrical about its pericentre direction. 
To explore what that pericentre direction might be, we found the angle at which the total integrated flux ratio (either side of an axis rotated around the stellar position) was equal to 1 (i.e., we found the total flux line of symmetry, accounting for the errors consistently as per our analysis of the fluxes in the two disc halves). 
This gave a value of $84.5{\pm}2.6^{\circ}$ clockwise from the minor axis (i.e., pointing between the SW and NW), shown on Fig.~\ref{fig:ALMAB7Tap} as the thick white dashed line. 
We also consider a combination of models \textbf{CL} and \textbf{ECC}, i.e., an eccentric disc which also has a clump of emission on the inner edge of the SW ansa (model \textbf{CL+ECC}).

\subsection{Scattered Light Analysis}
\label{sec:obsscatlight}
There are two main observational conclusions that we draw from Fig.~\ref{fig:HST_alone}. 
Firstly, along the major axis there is a radial offset (brightness asymmetry) in that the scattered light emission extends out to larger radii in the NE than the SW. 
To aid viewing this offset, the data has been annotated with dark-red dashed rings with radii of $5.7\arcsec$ and $6.7\arcsec$ (i.e., an average projected radii of ${\sim}108$\,au), which overlap with the +22\,mag\,arcsec$^{-2}$ surface brightness contour lines in the SW and NE ansae, and a black dash-dot line along the major axis. 
Similarly to the radial offset analysis in $\S$\ref{sec:radialProfAn}, we use these distances with an estimated error of ${\pm}0.1\arcsec$ to find a scattered light radial offset of $R_{\rm{scat}} {=} (r_{\rm{NE}}-r_{\rm{SW}}) / (r_{\rm{NE}}+r_{\rm{SW}}){=}0.081{\pm}0.011$. 
This demonstrates that there is a significant offset towards the NE direction in the scattered light emission in the outer region of the disc, reminiscent of those reported for the sub-mm emission in $\S$\ref{sec:radialProfAn} in the inner regions of the disc. 
This scattered light asymmetry is explored further in $\S$\ref{sec:discussion}.

Secondly, the scattered light emission along the minor axis in the SE region of the disc is brighter and more azimuthally extended in comparison to the NW. 
This can be seen by visually comparing the emission either side of the black dash-dot line (showing a position angle of $56^{\circ}$), which demonstrates there is more emission on the southern side than the northern. 
This type of asymmetry is usually inferred to be due to the dust emission preferentially forward scattering in this direction \citep{Augereau99}. 
Therefore this may suggest that the SE of the disc is on our near-side. 
However, since emission within ${\sim}2.25\arcsec$ cannot be probed due to coronagraph obscuration, this limits our ability to estimate the significance of this minor axis asymmetry. 
Higher resolution scattered light imaging that can observe closer to the star than 2.25$\arcsec$ is thus necessary to further constrain this.

\subsection{Summary of Observational Analysis}
\label{sec:SumObsAnalysis}
Taken in conjunction, we have found four significant observational constraints that we can place on the q$^1$~Eri debris disc, the first 3 from \textit{ALMA} (O1-3) and the fourth from \textit{HST} (O4), which will be used to constrain our modelling in $\S$\ref{sec:modelling} and discussed further in $\S$\ref{sec:discussion}: \\
\newline
\textbf{O1 - Broad Structure}: q$^1$~Eri has a bright, broad disc, with a position angle $\rm{PA}{=}57.0{\pm}1.0^{\circ}$, and an inclination $i{=}76.7{\pm}1.0^{\circ}$. 
This emission peaks at a radius of $81.6{\pm}0.5$\,au, and has a half-maximum width of $W_{\rm{HM}}{=}43{\pm}4$\,au, although likely extends from at least $34{-}134$\,au. 
The disc vertical scale height may have also been resolved with an aspect ratio $h{=}0.04{\pm}0.01$. 
\newline
\textbf{O2 - Major Axis Brightness Asymmetry}: the major axis emission brightness profile is asymmetric. 
At the same projected radii, the SW inner edge is brighter than the NE, but the outer edge is comparable on both sides. 
The inner edge asymmetry was quantified with the radial offset parameter $R_{\rm{inner}}{=}0.071{\pm}0.029$. 
The peak emission radius and outer edge were assessed similarly, finding values of $R_{\rm{peak}} {=} -0.001{\pm}0.006$ and $R_{\rm{outer}} {=} -0.026{\pm}0.033$ respectively. 
\newline
\textbf{O3 - Major Axis Flux Asymmetry}: the total Band~7 fluxes between $30{-}82$\,au and $30{-}200$\,au are greater on the SW side of the disc than on the NE side by fractional amounts $f_{\rm{82au}}{=}1.21{\pm}0.06$ and $f_{\rm{200au}}{=}1.23{\pm}0.04$ respectively. 
The Band~6 images are broadly consistent with the same ${\sim20}$ percent level of asymmetry as the Band~7 images, although the lower sensitivity at the longer wavelength precludes an independent determination. 
\newline
\textbf{O4 - Scattered Light}: the scattered light data shows the disc brightness to be radially and azimuthally asymmetric. 
We showed that dust emission has a significant radial offset towards the NE of the star for the 22\,mag\,arcsec$^{-2}$ emission contour lines, $R_{\rm{scat}}{=}0.081{\pm}0.011$, at an average projected radius of ${\sim}\rm{108\,au}$. 
We also found the SE side of the disc to be brighter than the NW, suggestive that the SE edge is on the disc's near-side to us. \\
\newline
Three possible scenarios were proposed in $\S$\ref{sec:obsInts} to model this system: model \textbf{CL}, a symmetric disc with an emission clump on the inner edge of the SW ansa; model \textbf{ECC}, an eccentric disc; and model \textbf{CL+ECC}, a combination of an eccentric disc with a clump on its SW inner edge. 
For comparison we will also present a symmetrical disc model \textbf{SYM}.

\subsection{CO J=3-2 Spectral Line Analysis}
\label{sec:discCO}
Separate to the continuum analysis, for the 2018 Band~7 \textit{ALMA} data, we produced a subset of the CASA measurement set with all spectral windows removed except the one containing the CO J=3-2 spectral line ($f{=}345.79599$\,GHz, herein referred to as $f_{\rm{CO}}$). 
Any circumstellar ${\rm{CO}}$ emission would be present across multiple data channels, as determined by the radial velocity of the star, $v_{\rm{rad}}{=}27.82{\pm}0.15\,\rm{kms^{-1}}$ \citep{Gaia18}, and the range of Keplerian velocities expected in the disc. 
To remove continuum emission from any $\rm{CO}$ signal, we used the CASA tool $\rm{uvcontsub}$ with a fit order of 1 external to the region defined by $f_{\rm{CO}} {\pm} \Delta f$, where $\Delta f$ was set to a velocity width ${\pm}35\,\rm{kms}^{-1}$, avoiding fitting to channels where $\rm{CO}$ could be present. 
We produced a data cube in the barycentric reference frame excluding all channels further than $200\,\rm{km\,s^{-1}}$ away from the J=3-2 line in velocity space using the CASA tclean algorithm (with zero clean iterations, since there was no significant emission per beam present in any channel). 

\begin{figure}
    \includegraphics[width=1.0\columnwidth]{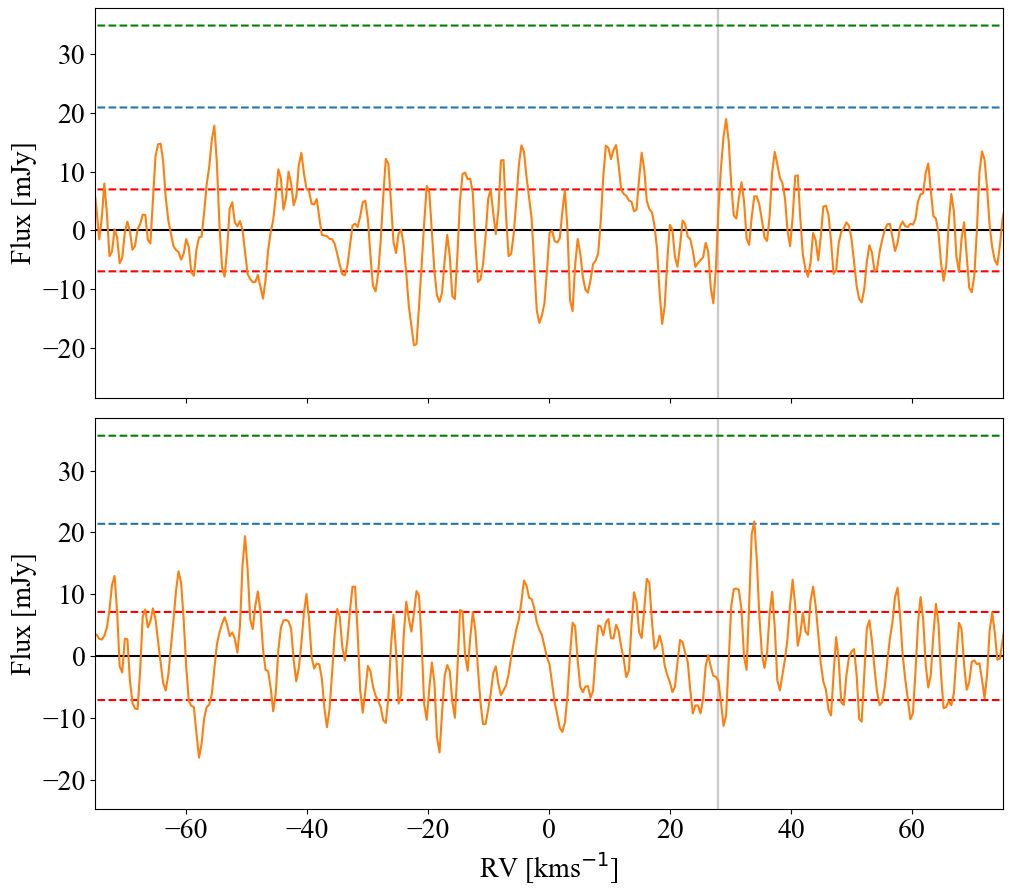}
    \caption{Flux as a function of radial velocity within spectrally and spatially filtered maps. The spectral shifts assume that the projected orbital motion is the clockwise and anti-clockwise directions in the top and bottom plots, respectively. The star's radial velocity is shown in grey.}
    \label{fig:fluxCOline}
\end{figure}

To measure the line profiles, we applied the spectro-spatial filtering method of \citet{Matra15,Matra17}. 
To do so, we applied a Keplerian mask to the data cube to shift channels in both of the possible disc rotation directions (clockwise and anti-clockwise) based on a stellar mass of 1.1\,$M_{\sun}$ \citep{Marmier13}. 
We integrated all emission in rotation directions between projected radii of ${\sim}$38-125\,au from the star (i.e., within the broad region defined by the full-width half-maxima of the sub-mm emission), which are both shown as a function of radial velocity in Fig.~\ref{fig:fluxCOline}. 
We found the rms of the two line profiles (clockwise and anti-clockwise rotations) across all velocities, and measured $\rm{rms{=}7.0\,\rm{mJy}}$ in the clockwise-rotated direction, and $\rm{rms{=}7.1\,\rm{mJy}}$ in the anti-clockwise rotated direction. 
We define our upper bound based on the larger of these two rms values, and place a $3\,\sigma$ limit on the CO line peak flux of $F_{\rm{CO,3\,\sigma}}{=}21.3\,\rm{mJy}$. 
The velocity channel spacing of the data is $\Delta v {=} 0.423\,\rm{kms}^{-1}$, however the effective spectral bandwidth is ${\sim}2.667$ larger than this, since adjacent \textit{ALMA} channels are not fully independent from each other\footnote{For a complete discussion of this, see \url{https://safe.nrao.edu/wiki/pub/Main/ALMAWindowFunctions/Note_on_Spectral_Response.pdf}}. 
The $3\,\sigma$ limit on this flux is then found as $\delta F_{\rm{CO},3\,\sigma} {=} 2.667 \times \Delta v \times \rm{rms} {=} 24.0\,\rm{mJy\,kms^{-1}}$. 
Based on this, we derive a gas mass upper bound in $\S$\ref{sec:discussionCO}.

\section{Modelling}
\label{sec:modelling}
Following the conclusion of the observational constraints in $\S$\ref{sec:SumObsAnalysis}, in this section we test these against parametric models of the Band~7 data (for simplicity), first outlining our modelling methodology, then giving best fit model parameters and discussing their implications.

\begin{table*}
    \centering
    \caption{Model best fit posterior values from models SYM, CL, ECC and CL+ECC. A dash indicates that a parameter was not a free parameter in the model this is associated with. In the case of $p_{\rm{inner}}$ the $3\,\sigma$ lower limit is shown along with the posterior value. The total Band~7 flux and peak emission radius, $F_{\rm{tot,B7}}$ and $r_0$, for each model is shown in the lowest section of this table (for the best fit model parameters), assuming a $10\%$ flux calibration error in $F_{\rm{tot,B7}}$ and $F_{\rm{star,B7}}$ (appropriate for interferometric observations with ALMA).}
    \begin{tabular}{l|c|c|c|c|c}
         \hline
         \hline
          && SYM & CL & ECC & CL+ECC \\
         \hline
         $n_{\rm{walkers}}$ && 120 & 180 & 140 & 180 \\
         $n_{\rm{steps}}$ && 600 & 1600 & 1000 & 2000 \\
         $n_{\rm{burn in}}$ && 250 & 1000 & 400 & 1200 \\
         \hline
         $M_{\rm{dust}}$ &[$M_{\oplus}$] & $0.0315{\pm}0.0008$ & $0.0275{\pm}0.0010$ & $0.0319{\pm}0.0008$ & $0.0282{\pm}0.0012$\\
         $r_{\rm{c}}$ &[au] & $75.5{\pm}1.1$ & $75.9{\pm}1.1$ & $75.1{\pm}1.1$ & $75.8{\pm}1.1$ \\
         $e$ && - & - & $0.053{\pm}0.009$ & $0.025{\pm}0.012$ \\
         $\omega$& [$^{\circ}$] & - & - & $11{\pm}15$ & $19{\pm}30$ \\
         $h$ && $0.054{\pm}0.004$ & $0.046{\pm}0.004$ & $0.055 {\pm} 0.004$ & $0.048{\pm}0.004$\\
         $i$ &[$^{\circ}$] & $78.62{\pm}0.20$ & $78.68{\pm}0.22$ & $78.58{\pm}0.20$ & $78.63{\pm}0.22$ \\
         $\rm{PA}$& [$^{\circ}$] & $56.67{\pm}0.20$ & $56.66{\pm}0.21$ & $56.65{\pm}0.21$ & $56.64{\pm}0.22$ \\
         $p_{\rm{inner}}$ && >5.1 ($16.8{\pm}3.9$) & >5.6 ($18.2{\pm}4.2$) & >5.5 ($18.1{\pm}4.2$) & >5.9 ($17.6{\pm}3.9$)\\
         $p_{\rm{outer}}$ && $-1.87{\pm}0.13$ & $-2.09{\pm}0.18$ & $-1.80{\pm}0.13$ & $-2.06{\pm}0.17$ \\
         $F_{\rm{star,B7}}$ &[$\rm{mJy}$] & $0.145{\pm}0.022$ & $0.142{\pm}0.022$ & $0.141{\pm}0.022$ & $0.144{\pm}0.022$ \\
         $x_{\rm{off,16}}$ &[$''$] & $0.057{\pm}0.021$ & $0.059{\pm}0.020$ & $0.044{\pm}0.022$ & $0.051{\pm}0.022$\\
         $y_{\rm{off,16}}$ &[$''$] & $-0.022{\pm}0.018$ & $-0.022{\pm}0.018$ & $-0.024{\pm}0.020$ & $-0.021{\pm}0.019$\\
         $x_{\rm{off,18}}$ &[$''$] & $0.004{\pm}0.028$ & $-0.024{\pm}0.028$ & $-0.11{\pm}0.03$ & $-0.03{\pm}0.04$ \\
         $y_{\rm{off,18}}$ &[$''$] & $-0.036{\pm}0.020$ & $-0.030{\pm}0.020$ & $-0.109{\pm}0.024$ & $-0.058{\pm}0.027$ \\
         \hline
         $F_{\rm{S}}$ &[$\rm{mJy}$] & - & $1.7{\pm}0.6$ & - & $1.7{\pm}0.6$ \\
         $x_{\rm{offset,S}}$ &[$''$] & - & $-4.2{\pm}0.5$ & - & $-4.3{\pm}0.5$ \\
         $y_{\rm{offset,S}}$ &[$''$] & - & $-2.6{\pm}0.3$ & - & $-2.7{\pm}0.4$ \\
         $R_{\rm{maj,S}}$ &[$''$] & - & $3.2{\pm}0.5$ & - & $3.5{\pm}0.6$ \\
         $R_{\rm{min,S}}$ &[$''$] & - & $2.3{\pm}0.5$ & - & $2.7{\pm}0.6$ \\
         $\rm{PA}_{\rm{S}}$& [$^{\circ}$] & - & $40{\pm}16$ & - & $50{\pm}50$ \\
         \hline
         $r_0$ &[au] & $82.9{\pm}1.2$ & $83.0{\pm}1.2$ & $82.9{\pm}1.2$ & $82.9{\pm}1.2$ \\
         $F_{\rm{tot,B7}}$ &[$\rm{mJy}$] & $11.3{\pm}1.1$ & $12.3{\pm}1.2$ & $11.9{\pm}1.2$ & $12.5{\pm}1.3$ \\
         \hline
    \end{tabular}
    \label{tab:bestFitVals}
\end{table*}

\subsection{Model Types and Definitions}
\label{sec:modTD}
The models described in $\S$\ref{sec:obsInts} are based on the same 12 parameter disc model. The 12 common parameters are the dust mass ($M_{\rm{dust}}$), characteristic radius ($r_{\rm{c}}$), scale height ($h$), inclination ($i$), position angle ($\rm{PA}$), inner and outer power-law indices ($p_{\rm{in}}$ and $p_{\rm{out}}$, described further below), photospheric Band~7 flux ($F_{\rm{star,B7}}$), and the phase centre offsets in RA and Dec for both Band~7 data sets ($x_{\rm{off,16}}$, $y_{\rm{off,16}}$, $x_{\rm{off,18}}$ and $y_{\rm{off,18}}$). 
This model can be configured to introduce (either or both) an extended 2D Gaussian source, and eccentricity. 
Where we include an additional extended source, this is defined by its flux ($F_{\rm{S}}$), offset from the stellar position ($x_{\rm{offset,S}}$ and $y_{\rm{offset,S}}$), a position angle ($\rm{PA}_{\rm{S}}$), and major and minor axis standard deviations ($R_{\rm{maj,S}}$ and $R_{\rm{min,S}}$), related to the FWHM via $\rm{FWHM}{=}2\sqrt{2\ln{2}}\,\sigma$. 

We parametrise the surface density ($\Sigma(a)$) of our model based on a two-component radial power law \citep{Augereau99, Kennedy18}, where on a grid of $r$ (distance from the star) and $\phi$ (azimuthal angle from the pericentre direction),
\begin{equation}
    \,\Sigma(a) \propto \Big[ \Big( \frac{a}{r_{\rm{c}}} \Big)^{-2\rm{p_{in}}} + \Big( \frac{a}{r_{\rm{c}}} \Big)^{-2\rm{p_{out}}} \Big]^{-1/2},
    \label{eq:powerlaw}
\end{equation}
for $a{=}r[1 + e \rm{cos}(\phi-\omega)]/(1 - e^{2})$, where $\omega$ is the angle between the pericentre direction and the line of nodes where the disc plane crosses the sky plane (i.e., the major axis, for which $\omega{=}0^{\circ}$ is the south-western direction). 
This parametrisation of the surface density is valid for a disc containing particles with the same eccentricity and pericentre \citep{Marino19}; note, however, that it is possible that particle eccentricities and orientations are not all identical in such a broad disc, and we discuss the implications of this in $\S$\ref{sec:modelSum}. 
The models have the same vertical Gaussian density distribution, defined by the scale height, $h{=}H/r$, where $H$ is the height of dust above the midplane at a radius $r$ \citep[e.g.,][]{Marino16, Marino19}. 
Where used, eccentricity $e$ is a constant. 
All models use the same minimum and maximum grain sizes of $a_{\rm{min}}{=}0.9\,\mu \rm{m}$ and $a_{\rm{max}}{=}1$ $\rm{cm}$ (respectively set by the grain blow out size, and by neglecting emission from larger grains at Band~7 wavelengths), a dust grain density of $2.7$\,$\rm{g}$\,$\rm{cm}^{-3}$, a grain-size distribution with power-law exponent $\alpha{=}3.5$ \citep{Dohnanyi69}, and a weighted mean dust opacity based on a mix of compositions with mass fractions of 70\% astrosilicates, 15\% amorphous carbon and 15\% water ice \citep[for example, identical to that used to model HD107146 by][i.e., $1.88\,\rm{cm}^2\,\rm{g}^{-1}$ at $\lambda{=}856\,\mu$m]{Marino18}. 
For model CL (described in detail in $\S$\ref{sec:modFitResComp} we show the dust density and temperature profiles in Appendix~\ref{sec:appendixDust}, which shows a temperature of T${=}44$K at the location where the dust density peaks.

The Band~7 emission from the star was estimated to be ${\sim} 99$\,$\mu$Jy in \S \ref{sec:SED}. 
However, since stellar emission at sub-mm wavelengths is often poorly constrained, the unresolved flux at the stellar location in the Band~7 image is left as a free parameter $F_{\rm{star,B7}}$.

\begin{figure*}
    \includegraphics[width=2.0\columnwidth]{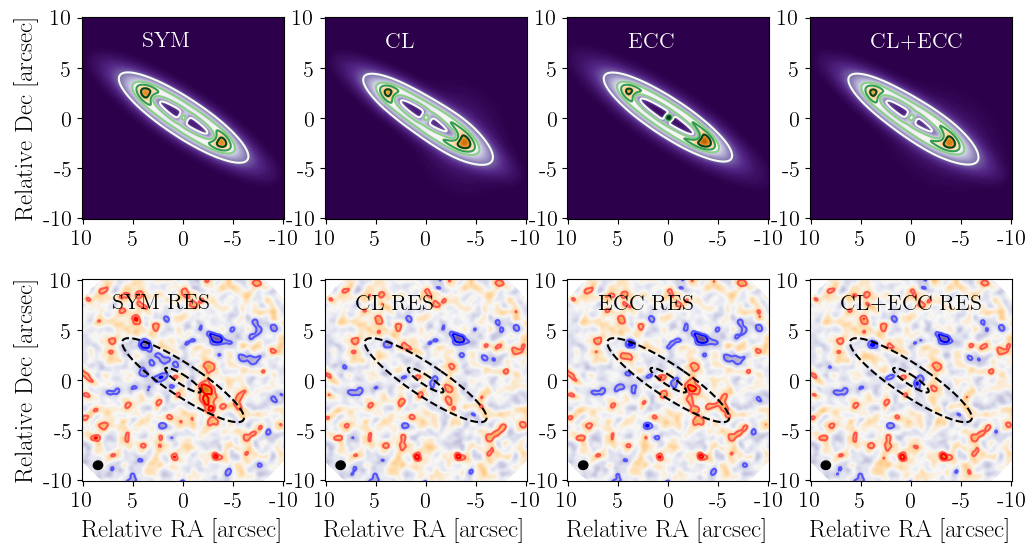}
    \caption{Plots to show the best fit parameter models (top) and the residual maps produced when the models are subtracted from the concatenated Band~7 data sets (bottom) for model SYM (left), model CL (centre-left), model ECC (centre-right) and model CL+ECC (right). Model contours were chosen to demonstrate their inner and outer edges and how their emission varies in the two ansae. On the residual maps the black dashed ovals demonstrate the projected disc emission at the half-maximum inner and outer edges, and the contours demonstrate the ${\pm}2\,\sigma$ and ${\pm}3\,\sigma$ residuals (red positive, blue negative). On the lower left is the beam size for each of these images.}
    \label{fig:ModelResMap}
\end{figure*}

\subsection{Model Fitting, Results and Comparison}
\label{sec:modFitResComp}
We used the $\rm{RADMC-3D}$ \citep{Dullemond12} package to compute the dust temperature of the disc and produce model images defined by the parameters outlined in $\S$\ref{sec:modTD}, at a Band~7 wavelength of $856\mu \rm{m}$, and compute model visibilities at the same uv-baselines as our \textit{ALMA} observations using the tools developed in \citet{Marino18}.
All models are produced with the same number of visibility data points as the combined 2016 and 2018 Band~7 observations, $N_{\rm{dat}}{=}2\times N_{\rm{Vis}} {=} 19,440,912$ (9,720,456 for both real and imaginary components). To investigate the consistency of these models with our Band~7 data, we used the MCMC best fit model-parameter estimator, $\rm{emcee}$ \citep{Goodman10, FM13}. The posterior distribution generated by $\rm{emcee}$ is the product of the prior distribution and the likelihood function, which takes a value $\rm{exp}(- \chi ^{2}/2)$, where for interferometric data sets
\begin{equation}
    \chi^2_{\rm{Vis}} = \sum_{i=1}^{\rm{N_{vis}}} \frac{|| V_{\rm{data,i}} - V_{\rm{model,i}} ||^2}{f_{\,\sigma}^2\delta V_{\rm{data,i}}^2},
    \label{eq:chi2}
\end{equation}
in which $V$ is the value of the each visibility, $\delta V$ is the intrinsic dispersion of the visibilities (as calculated by the CASA package $\rm{statwt}$), and $f_{\,\sigma}$ a normalisation parameter discussed further below. 

For large interferometric data sets, individual visibility measurements can be dominated by noise (i.e., $S/N<<1$), leading to the reduced $\chi^2_{\rm{Vis}}$ values ($\chi_{\rm{red}}^2{=} \chi^2_{\rm{Vis}}/\rm{N_{dat}}$) of even null models being close to 1 (i.e., the value expected for a perfect model). 
From previous \textit{ALMA} modelling of debris discs, it has been shown that the CASA routine $\rm{statwt}$ (used to re-weight visibilities according to their dispersion) does not provide an accurate measure of the absolute uncertainties on visibility measurements, although it does provide a good estimate of the relative errors between visibilities. 
This means that the weights or variance can be off by a small factor, and subsequently affect the posterior distributions of modelled parameters (i.e., leading to incorrect parameter estimations and errors). 
To account for this, as per previous analyses of \textit{ALMA} data \citep[e.g.,][]{Marino18}, we apply a correction factor, $f_{\,\sigma}$ (see Equation~\ref{eq:chi2}), such that $\chi_{\rm{red, null}}^2{=}1$ (i.e., for a model with $V_{\rm{model}}{=}0$ everywhere). 
For both 2016 and 2018 data sets we find a similar $f_{\,\sigma}\sim1.6$ (i.e., this confirms that $\rm{statwt}$ has underestimated the uncertainty in the visibilities). 
The need to scale uncertainties by $f_{\,\sigma}$ however means that the absolute value of $\chi^2$ cannot be used to assess the goodness-of-fit of an individual model (e.g., since the null model should have $\chi_{\rm{red}}^2>1$ as it should be a poor fit). 
However, given that $f_{\,\sigma}$ is constant, the $\chi_{\rm{Vis}}^2$ values can still be used to quantify the improvement between different models. 
We outline our assessment of the goodness-of-fit in the following section.

\subsubsection{Model Results}
\label{sec:modRes}
We investigated the consistency of our data with a symmetrical model \textbf{SYM}, defined by the 12 common parameters introduced in $\S$\ref{sec:modTD}, before continuing with models \textbf{ECC}, \textbf{CL} and \textbf{CL+ECC}, with 14, 18 and 20 free parameters respectively. 
For these MCMC routines, between 140-200 walkers and 600-2000 steps were used to ensure parameter solutions were converged on, with burn in lengths varying between 250 and 1200. 
The results of our modelling procedure are shown in Table~\ref{tab:bestFitVals}, where we report all best fit model values, along with $r_0$ (the average radius of the peak brightness found from model radial profiles) and the \textit{total} Band~7 flux, $F_{\rm{tot,B7}}$ (which includes $F_{\rm{star,B7}}$), generated by each best fit model, for which we assume and quote a 10\% flux calibration error.

\begin{table*}
    \centering
    \caption{Model comparison table, demonstrating in the upper section of the table the $\chi^2$ values from the visibilities and images, and the difference in the visibility $\chi^2$ values (from model SYM), and then results for observational conclusions O1 (half-peak disc width, noting that all other O1 conclusions are readily generated by the models and are included in Table~\ref{tab:bestFitVals}), O2 Maj (major axis radial offsets for the major axis) and O3 Maj (the flux ratio for emission compared in the major axis between $30-82$\,au and between $30-200$\,au) in the following table sections. Note that all values in red and bold are $>3\,\sigma$ from the equivalent metric determined in the Observational Analysis (or in the case of the $\chi_{\rm{Im}}^2$ metric, a poor representation of the data), and those values in blue are between $2{-}3\,\sigma$ from the same equivalent observational metric. For ease of comparison, the observational analysis parameters are shown in the right-most column.}
    \begin{tabular}{l|c|c|c|c|c|c}
         \hline
         \hline
          && SYM & CL & ECC & CL+ECC & Obs Analysis \\
         \hline
         $\chi_{\nu,\rm{Im}}^2$ && \textbf{\color{red}{2.77}} & $1.13$ & \textbf{\color{red}{1.72}} & $1.13$ & - \\
         $\chi_{\rm{Vis}}^2$ && $19525531.6$ & $19525462.8$ & $19525499.4$ & $19525457.0$ & - \\
         $\Delta \chi_{\rm{Vis}}^2$ && $0.0$ & $-68.8$ & $-30.2$ & $-74.6$ & - \\
         $\Delta \rm{BIC}_{\rm{Vis}}$ && $0.0$ & $31.9$ & $3.4$ & $59.7$ & - \\
         $\Delta \rm{BIC}_{\rm{Im}}$ && $0.0$ & $-191.8$ & $-103.2$ & $-181.2$ & - \\
         $N_{\rm{par}}$ && 12 & 18 & 14 & 20 &  - \\
         \hline
         \textbf{O1} (Width) && & & & \\
         $W_{\rm{HM}}$& [au] & \color{blue}{54} & \color{blue}{55} & 50 & \color{blue}{54} & $43{\pm}4$ \\
         \hline
         \textbf{O2 Maj} && & & & \\
         $R_{\rm{peak}}$ && $0.00$ & $0.00$ & \textbf{\color{red}{0.049}} & $0.00$ & $-0.001{\pm}0.006$ \\
         $R_{\rm{inner}}$ && \color{blue}{0.00} & $0.048$ & $0.068$ & 0.040 & $0.071{\pm}0.029$ \\
         $R_{\rm{outer}}$ && $0.00$ & $0.00$ & 0.013 & $0.00$ & $-0.026{\pm}0.033$ \\
         \hline
         \textbf{O3 Maj} & && & & \\
         $f_{\rm{82\,au}}$ && \textbf{\color{red}{1.00}} & 1.30 & \color{blue}{1.39} & 1.27 & $1.21{\pm}0.06$ \\
         $f_{\rm{200\,au}}$ && \textbf{\color{red}{1.00}} & 1.17 & \textbf{\color{red}{0.98}} & 1.14 & $1.23{\pm}0.04$ \\
         \hline
    \end{tabular}
    \label{tab:modelComparison}
\end{table*}

We first discuss the ability of all four models to explain the axisymmetric parameters, before their ability to interpret the asymmetric parameters. 
The average peak emission radius $r_0$ for all models is found to be within ${<}1\,\sigma$ from the value determined in the observational analysis. 
Likewise comparison of best fit parameters with those derived directly from the cleaned images shows the total Band~7 flux is within $1.3\,\sigma$, the inclination within $1.9\,\sigma$, the position angle within $0.4\,\sigma$, the stellar flux within $1.5\,\sigma$, and all stellar offsets consistent with 0 (no offset), within $3\,\sigma$ (for models SYM, CL and CL+ECC). 
In the case of model ECC, the offset of the 2018 data is not found at the origin (albeit nearby). 
These four models do however show broad agreement with these aspects. 
We did not derive estimates for the dust mass, or outer radial power law indices in the observational analysis, however these parameters are relatively well constrained by the four models. 
The inner edge power law index is however only poorly constrained, and the values presented in Table~\ref{tab:bestFitVals} show the $3\,\sigma$ lower limit, and the MCMC-derived best-fit value and error. 
Although only weakly constrained, in all models the inner edge index is shown to be sharp, and the better constrained outer edge index is shown to be shallow, which both appear consistent with the radial profiles in Figs.~\ref{fig:ALMAB7RadProfs}. 
That all models find a scale height significantly above 0 is discussed in $\S$\ref{sec:discussion}, however this suggests that these models have at least partially resolved the vertical distribution of the disc's emission. 
In $\S$\ref{sec:radialProfAn} we showed that this was consistent with $h{\sim}0.04$, and thus these values are all broadly in agreement with this (i.e., are at the few percent level). 

Whilst the broad disc parameters can be well modelled by each of the 4 models, we next assess which of these provides the best quantitative and qualitative fit. 
In Fig.~\ref{fig:ModelResMap} we show images of all four models (top) and the 2016 and 2018 epoch concatenated residual maps, re-imaged using $\rm{tclean}$ with a $\rm{uvtaper}$ of $0.5''$ (to match the image of Fig.~\ref{fig:ALMAB7Tap}) for each of the four models (bottom). 
Qualitatively we can see that whilst these models are similar, there are key differences in how these reproduce the overall disc morphology, and the relative emission brightness in different regions of the disc. 
In the residual maps, models SYM and ECC have multiple regions coincident with the disc where the data is fitted poorly (residuals ${>}3\,\sigma$). 
In model CL+ECC there is one region on the outer edge of the NE ansa where the residual map exceeds $3\,\sigma$ (and a few $2\,\sigma$ contours), whereas the residual map of model CL appears to fit the data most consistently (all residuals ${<}3\,\sigma$, with only a few $2\,\sigma$ contours). 
We note that despite model CL+ECC having 2 more free parameters than model CL, yet model CL appearing to be a qualitatively better fit, this difference is consistent with our quantitative assessment of the $\chi^2$ values (discussed later in this section).

To quantify this, we report in Table~\ref{tab:modelComparison} the parameter $\chi_{\nu,\rm{Im}}^2$ as the reduced $\chi^2$ of the residual maps within projected radii $38{-}125$\,au (based on the peak emission radius $\pm$ the half-peak emission width, see Table~\ref{tab:radProfVals}) such that a significant extent of the inner and outer edge emission was included. 
This required us to define the residual map error, which we determined from the rms of the residual maps external to the region that $\chi_{\rm{Im}}^2$ is calculated within as $\rm{15.9\mu Jy\,beam^{-1}}$. 
Consistent with our qualitative assessment, the values shown in Table~\ref{tab:modelComparison} demonstrate that models CL and CL+ECC are most consistent with a value of 1.0 (1.13 for both), and models ECC and SYM are both higher (1.72 and 2.77 respectively). 

To assess how well these compare with respect to the mean and uncertainty in these residual maps, we start by noting that a $\chi_{\rm{Im}}^2$ value of 1.0 would imply that the residual maps are consistent with noise. 
By producing $N{=}3000$ maps of noise convolved with a beam equal to that in our residual maps (with a resulting rms equal to $\rm{15.9\,\mu Jy\,beam^{-1}}$), we found $\chi_{\rm{Im}}^2$ exceeded 1.13 in ${\sim}4\%$ of all iterations, and in no circumstances were values found ${>}1.72$, which based on the number of runs, allowed us to place an upper bound on the frequency at which we would expect residual maps to exceed this as ${<}0.03\%$. 
We found these results were normally distributed (with a slight tail towards higher values), centred on 0.96, with a width of 0.10. 
We therefore assess the CL and CL+ECC residual maps as being consistent with a map comprised entirely of noise (i.e., given their $\chi_{\rm{Im}}^2$ values are both 1.13), but that both the ECC and SYM residual maps are inconsistent. 
This implies from the $\chi_{\rm{Im}}^2$ values that models CL and CL+ECC are both reasonable representations of our data, and models SYM and ECC are poor representations. 

We next compare the visibilities from their entire interferometric data sets respectively (i.e., rather than between projected radii), and report the $\chi_{\rm{Vis}}^2$ values for each of the models in Table~\ref{tab:modelComparison} with their respective differences, $\Delta \chi_{\rm{Vis}}^2$, from the value achieved with model SYM. 
Subtracted as such, this would mean that the most negative $\Delta \chi_{\rm{Vis}}^2$ would represent the best fitting model, i.e., in order of best-to-worst fit, this would conclude that the models are ordered CL+ECC, CL, ECC and finally SYM. 
Since the difference between model CL and model CL+ECC is $\Delta \chi_{\rm{Vis}}^2{\sim}6.8$, these two models provide a comparable fit to the data, whereas models ECC and SYM are correspondingly worse fits. 
Thus we anticipate that the two additional free parameters required for model CL+ECC are not justified, and so that model CL is favoured by this analysis.

We further this assessment, by calculating the Bayesian Information Criterion (BIC), which statistically assesses the significance of additional free parameters, where
\begin{equation}
    \rm{BIC} = \chi^2 + N_{\rm{par}}\ln{N_{\rm{dat}}},
    \label{eq:BIC}
\end{equation}
for a model with $\rm{N_{par}}$ independent free parameters, and $\rm{N_{dat}}$ independent data points. 
This is reported in Table~\ref{tab:modelComparison} for each of the models relative to the baseline of model SYM, for consideration of both the visibilities and the image, i.e., the $\Delta \rm{BIC_{Vis}}$ and $\Delta \rm{BIC_{Im}}$ values respectively. 
Where the $\Delta \rm{BIC}$ between two models exceeds 10, this is often reported as statistically strong evidence to support the hypothesis that the model reporting the \textit{lower} BIC is a more significant fit to the data than the model it is being compared with \citep{Schwarz78,Kass95}. 

This analysis however arrives at a different conclusion, with the model SYM being statistically favourable, with the model ECC being broadly comparable, but CL and CL+ECC being consecutively worse. 
This is surprising given that the residual image of model SYM shows that this is a poor fit to the observations, and that models CL and CL+ECC are far better. 
Thus whilst we may use the $\rm{BIC}$ parameter to distinguish between these models, it may not provide a robust metric to do so. 
To understand why this may have occurred, consider that the BIC parameter scales with $\rm{ln(N_{Vis})}$, and so, for the number of visibility measurements in our data set, \textit{each} additional model parameter would require a relative improvement in $\chi_{\rm{Vis}}^2$ of ${>}16.8$ to be justified. 
However, our data also contain a high fraction of data points that are negligibly affected by the model or the asymmetry, resulting in the BIC applying a disproportionate penalty to additional model parameters. 
In the image space there are orders of magnitude fewer independent data points, therefore this issue is much less pronounced. 
The $\Delta \rm{BIC_{Im}}$ values support this, finding that the model CL is a significantly better fit than any other models (with a value ${>}10$ lower than the next statistically favoured model CL+ECC), in line with our earlier assessment of the residual image. 
Nevertheless, we acknowledge that the conclusion from the BIC analysis may also reflect that our asymmetric models are also not a perfect representation of the data.

The lower three sections of Table~\ref{tab:modelComparison} assess how well the model images reproduce the observationally derived parameters associated with points O1, O2 and O3 from $\S$\ref{sec:SumObsAnalysis} (when convolved with the same beam), noting that point O4 is beyond the scope of the sub-mm models considered here. 
As previously discussed, all models are consistent with the broad structure of the disc noted in point O1, such as the peak radius, inclination and position angle. 
Table~\ref{tab:modelComparison} shows that all 4 models reproduced the disc width to within $3\,\sigma$, consistent with the Band~7 image (although all model discs were wider than the Band~7 image, measured as 43\,au). 
We also note that the Band~7 SW-NE flux difference $\delta F{=}1.33{\pm}0.18$\,mJy is strongly consistent with the clump flux value, $F_{\rm{S}}{=}1.7{\pm}0.6$\,mJy (i.e., within $1\,\sigma$ for both clump models). 
Only models CL and CL+ECC are able to account for the major axis offsets within $2\,\sigma$, however, this is not a strong conclusion given that the values for the model SYM (i.e., without any offset) would also be deemed reasonable (i.e., all within $3\,\sigma$), and since the clump location is not tightly constrained on the SW ansa (see Appendix~\ref{sec:appendixA}). 
Further, only models CL and CL+ECC are able to reproduce the major axis flux ratios $f_{\rm{82\,au}}$ and $f_{\rm{200\,au}}$ within $2\,\sigma$, with both model ECC and SYM providing poorer estimates of these parameters. 
This comparative analysis thus suggests that despite a few shortcomings, the CL and CL+ECC asymmetric models have been able to model the major axis asymmetries reasonably well, with the ECC model proving less adequate, though still preferable to the symmetric model.

\subsection{Modelling Summary}
\label{sec:modelSum}
Of our four models, two are able to reproduce the observations to within the noise levels (see observational parameters in Tables~\ref{tab:bestFitVals} and~\ref{tab:modelComparison}), and consistency was shown with all the broad disc parameters and our measurements in $\S$\ref{sec:obsAnalysis} in all four cases. 
The asymmetric parameter analysis demonstrated that the models which introduced a clump on the SW ansa of the debris disc were favourable, and by further analysing the respective residual images and $\chi^2$ values from the model visibilities we find that the model with a single clump is the best fit to our data. 
Although, we note that an assessment of the $\rm{BIC}$ parameters may suggest differently to the above, with the models with fewest parameters being shown to be consecutively more favourable by this metric. 
Nevertheless, in conjunction we consider our model that is a symmetrical disc with a clump model (model CL) to be the most favourable scenario of those considered. 

We note however that the clump modelled was a simple 2D Gaussian. 
A more complicated clump profile may better model this system with more success given that this distribution may be too simplistic to interpret the physical origin of the extra emission (i.e., this may be azimuthally extended around an orbit if the emission is due to a collision within the belt, which we consider in $\S$\ref{sec:discCollision}). 
On the other hand, whilst the disc models without a clump (either fully symmetrical or having an underlying constant eccentricity distribution) were poorer fits to the data, an eccentric model with both free and forced eccentricity terms, or one with a outwardly falling eccentricity (e.g., if due to the forced eccentricity of a planet sculpting the inner edge) may provide a better fit than the constant eccentricity modelled here \citep[for example, similar to the models of][]{MacGregor17, Kennedy20}. 
We suggest future work may wish to conduct detailed modelling of this scenario to explore the origin of the asymmetric emission of this debris disc, which we further discuss in $\S$\ref{sec:discOriginAsymms}. 
Thus, whilst our single clump model provides a reasonable model of our data, we leave open the possibility that more complicated clump profiles or multi-component eccentric distributions may better and consistently interpret our observations.

\section{Discussion}
\label{sec:discussion}
In $\S$\ref{sec:obsAnalysis} we presented an analysis of the Band~6 and 7 images for q$^1$~Eri. From this we were prompted to investigate the disc in Band~7, given the observed asymmetries, and presented different models for the disc morphologies in $\S$\ref{sec:modelling}. Here we interpret these results and discuss them in the context of the wider q$^1$~Eri planetary system.

\subsection{A Massive and Broad Debris Disc}
\subsubsection{How big are the largest planetesimals, and what does this imply for the disc mass?}
\label{sec:discMASS}
As outlined in $\S$\ref{sec:intro}, q$^1$~Eri stands out as having the brightest debris disc of the closest 300 sun-like stars, despite its age being $1.4{\pm}0.9\,\rm{Gyr}$ \citep{Marmier13}. 
This could mean that q$^1$~Eri's disc is somehow extreme among the population \citep[e.g., in having started with an unusually large mass in planetesimals, see][]{Kral15}, or alternatively that there is more dust in this system than expected from steady state collisional models (e.g., due to a recent transient event). 
The disc in this system is broad, extending between $34{-}134$\,au with a peak at ${\sim}81.6$\,au. 
Despite being broad, q$^1$~Eri's peak emission radius in the sub-mm is in accordance with the population of 26 \textit{SMA} (Sub-Millimetre Array) and \textit{ALMA} resolved discs in \citet{Matra18}, which (for $L_{*}{\sim}1.59L_{\odot}$) predicts a sub-mm disc radius $r_0{=}73{\pm}12$\,au (i.e., within $1\,\sigma$ of our measurement). 
Further, this sub-mm resolved radius is entirely consistent with the archival far-IR \textit{Herschel} radius of $R_{\rm{disc}}{=}81.1^{+1.8}_{-1.3}$\,au \citep{Marshall21}. 
Whilst our measurement of the disc width is however narrower than the broad $\Delta R_{\rm{disc}}{=}71.1^{+1.9}_{-13.3}$\,au value found by \citet{Marshall21}, this is likely due to the different manner in which we assess the width (i.e., we note that the extent of the disc is much broader than the ${\sim}43$\,au width we define from the half-maximum emission).
Further, the width measurements of \citet{Marshall21} are also likely unresolved, which for example, may also be influenced by emission from an inner warm component that we discuss further in $\S$\ref{sec:discussionWarm}. 

To constrain the disc's total mass, we first quantify the size of planetesimals that must be feeding the disc's collisional cascade, assuming this to be in steady state. 
To do so we assume the number of bodies with a given size follows a power law with slope $\alpha{=}-3.5$ \citep{Dohnanyi69} from dust with a maximum modelled size of $D{\sim}2\,\rm{cm}$, in which there is a mass $M_{\rm{dust}}{=}0.028\,M_{\oplus}$ (determined from model CL), up to a largest planetesimal size $D_{\rm max}$, which results in a total mass of $(D_{\rm max}/D_{\rm mm})^{0.5}M_{\rm{dust}}$. 
The largest planetesimal size is determined by setting the collisional lifetime $t_{\rm{coll}}$ \citep[see eq.~3 of][]{Wyatt08} to be equal to the age of $1.4\,\rm{Gyr}$ (although we note this age is uncertain at the ${\sim}50\%$ level). 
This gives a lower limit largest planetesimal size of $D_{\rm{max}}{\sim}1\,\rm{km}$ and a lower-bound total disc mass estimate of $M_{\rm{disc}}{>}8\,M_{\oplus}$. 
These are lower limits because there may be even larger planetesimals in the disc, which have not yet contributed to the collisional cascade, and so are not required to be present \citep[e.g.,][]{Wyatt02}. 
These values are found for a fixed $Q_{\rm{D}}^{*}{\sim}163\,\rm{J\,kg}^{-1}$, consistent with ${\sim}1\,\rm{km}$ Basalt planetesimals for both \citet{Benz99} and \citet{Stewert09}, a relative velocity $v_{\rm{rel}}{=}103.9M_\star^{0.5}\,h\,r^{-0.5}{=}620{\pm}30\,\rm{m\,s}^{-1}$ \citep[using equation 10 of][with our model CL determined value for $h$]{Matra19}, the stellar mass $M_{*}{\sim}1.1\,M_{\odot}$, at the peak brightness radius $r{=}81.6{\pm}0.5$\,au, and with a belt width $dr{\sim}43$\,au. 
Given the uncertainties and assumptions (e.g., constant power law slope, emission being dominated by grains with diameters smaller than 2\,cm), these lower limits are only approximate, but they are consistent with the ${\sim}81.7\,M_{\oplus}$ disc mass determined by the detailed collisional modelling of \citet{Schuppler16}. 
Indeed, given that this collisional modelling included planetesimals up to 100\,km in size and resulted in a disc mass a factor of 10 higher than our ${\sim}8\,M_\oplus$ lower bound, the relationship between total disc mass and maximum planetesimal size (i.e., $M_{\rm{disc}}{\sim}D_{\rm{max}}^{0.5}$) suggests these two estimates may be strongly consistent. 
This shows that, while the required disc mass is certainly large, it is not unreasonably so, suggesting that the disc's unusual brightness can still be explained within the context of steady state collisional erosion. 

Although we have thus far assumed that planetesimals are stirred at $t{=}0$ (i.e., born stirred), it is also possible that planetesimal collisions were delayed until a time $t_{\rm{delayed}}$ in accordance models \citep[e.g., self-stirring or planet-stirring][]{Kenyon04b, Mustill09}. 
Such delayed stirring may commence between 10s of Myr to Gyr timescales, and thus only recently in the q$^1$~Eri debris disc. 
Figure 5 of \citet{Wyatt08} shows the fractional luminosity evolution with one such delayed stirring model for a planetesimal belt extending from 30-150\,au (i.e., similar to the extent measured here in the sub-mm for q$^1$~Eri). 
If this belt is stirred after just ${\sim}20$\,Myr, then by ${\sim}1.4\,$Gyr of evolution this model finds $f{\sim}3{\times}10^{-4}$ (consistent with the broad planetesimal belt of q$^1$~Eri, $f{=}$($2.44{\pm}0.08$)${\times}10^{-4}$, see $\S$\ref{sec:SED}) for a disc with an initial mass of one minimum mass solar nebula (1.0\,MMSN). 
Whilst the delayed self-stirring models initiate the collisional cascade after the formation of 2000\,km planetesimals within the disc, planets outside the disc can initiate the cascade in planet-stirring models. 
This suggests that belts as bright and old as q$^1$~Eri may only contain as much mass as the MMSN, which seems reasonable since at least this mass must have been present in the Solar System. 
Given the range of initial disc masses, extents and plausible range of delayed stirring timescales that may be consistent with the observed fractional luminosity of q$^1$ Eri, the possibility that the disc was not stirred at $t{=}0$ adds uncertainty to any disc mass calculations. 
We consider the mass of planetesimals in the belt further in $\S$\ref{sec:discvertscal}. 

\subsubsection{Are there further constraints from the vertical distribution?}
\label{sec:discvertscal}
The scale height of the disc is resolved in all of the models presented here, with an average value ${\sim}0.05$, comparable with that determined for other debris discs of $0.02{-}0.12$ \citep{Hughes18, Kennedy18, Daley19, Marino19, Matra19}. 
This agrees with both the observational and modelling conclusions that we may have resolved the vertical dust distribution. 
The vertical distribution in discs can be interpreted as due to dynamical stirring from massive planetesimals. 
Equating the relative velocity, $v_{\rm{rel}}$ (as estimated in $\S$\ref{sec:discMASS} at the peak emission radius ${\sim}81.6$\,au), to the escape velocity of the largest bodies \citep[e.g.,][]{Daley19}, for an asteroid density of $\rho{=}2.7\,\rm{g\,cm^{-3}}$, we find lower bounds on the mass and size of the bodies providing the stirring as $M_{\rm{body}}{>}2.3{\pm}0.5{\times}10^{21}\,\rm{kg}$ and $D_{\rm{max}}{>}1200\,\rm{km}$, respectively ${\sim}20\%$ the mass and ${\sim}50\%$ the size of Pluto. 
From equation 12 of \citet{Matra19}, \citep[based on][]{Ida93}, our model determined value of $h$, and $M_{\rm{body}}$, we derive a surface mass density  $\Sigma{=}0.039\,M_\oplus\,$au$^{-2}$. 
Given this value of $\Sigma$ is around 40 times larger than the MMSN at 81.6\,au, this suggests that bodies larger than 1200\,km indeed may be necessary in order to stir the disc within the age of the system (i.e., to reduce the surface mass density to levels that may be more plausible). 

This dynamical limit on planetesimal sizes could suggest that these may exist in the belt a factor of $10^3$ larger than the ${\sim}1\,$km sized lower limit predicted by collisional replenishment. 
However, if the assumed $\alpha{=}3.5$ \citet{Dohnanyi69} size distribution continued up to planetesimals of this size, then the disc mass would be $M_{\rm{disc}}{\sim}220\,M_{\oplus}$, a factor of ${\sim}3$ higher than the prediction of \citet{Schuppler16} and over an order of magnitude larger than the lower limit derived from the collisional lifetime and age of the system. 
This would not violate our earlier calculation since these ${\gg}1$\,km planetesimals could be abundant in the disc without having collided within the age of the system. 
Whilst this higher total disc mass is still consistent with the dust mass measurements of protoplanetary discs \citep[for example, see][]{Andrews05, Ansdell16, Cieza19}, high debris disc masses, i.e., those in the range of $100{-}1000\,M_{\oplus}$, become problematic since these would require a very high efficiency of primordial dust being incorporated into these larger planetesimals \citep[an example of the so-called 'disc mass problem', see][]{Krivov18, Krivov21}. 
Nevertheless, it might still be possible to explain the observed level of stirring by embedded bodies while retaining a lower disc mass (i.e., if the largest bodies are less frequent than the $\alpha{=}3.5$ size distribution would predict). 
If instead the size distribution had a slope of $\alpha{=}3.7$, then even with these 1200\,km bodies, the total disc mass estimate would be reduced from ${\sim}220\,M_\oplus$ to ${\sim}6\,M_\oplus$, since the total number of these would be greatly reduced. 
On the other hand, the size distribution could be truncated, being much steeper for planetesimals larger than a few kilometers, and shallower in other regions. 
We estimate the effect that this can have on the derived disc mass using equation 9 of \citet{Krivov21}, and find that the total mass can be reduced by an order of magnitude from ${\sim}220\,M_\oplus$ to ${\sim}17\,M_\oplus$, if based on a triple power law size distribution, with $q_{\rm{med}}{=}4$, $q_{\rm{big}}{=}3$ and $q{=}3.5$. 
Alternatively the measured vertical scale height could be due to other dynamical interactions (e.g., stirring by a planet internal or external to the belt, or a recent stellar fly-by) or even be a remnant of the primordial disc \citep[e.g., if this disc was born stirred, ][]{BoothR16}. 
In summary, whilst q$^1$~Eri at an age of ${\sim}1.4\,\rm{Gyr}$ is an outlier in terms of its brightness, it need not be an outlier in terms of its disc mass unless many much larger planetesimals are present.

\subsubsection{What is happening at the disc inner-edge?}
\label{sec:discIE}
Although this disc is radially broad, it also has a sharp inner edge. 
To within $3\,\sigma$, model CL has an inner edge $\,\Sigma (r)\propto r^{>5.6}$ (see Table~\ref{tab:bestFitVals}). 
\citet{Kennedy10} showed that disc inner edges shaped by collisional processes have much shallower inner edges of $\,\Sigma (r) \propto r^{7/3}$. 
That our models find a steeper inner edge favours a planetary carving scenario \citep[see, e.g.,][]{Chiang09}, though we cannot rule out that a steep edge simply reflects the primordial planetesimal distribution. 

Already known in this system is the exo-Jupiter, q$^1$~Eri~b, but with an orbital radius at ${\sim}2$\,au this planet cannot be responsible for carving the inner edge (which we place at ${\sim}57$\,au based on the half-maximum derived inner edge from Table~\ref{tab:radProfVals}). 
The lack of any gaps in the disc beyond ${\sim}57$\,au suggests there is no evidence for additional ${\sim}$planet-mass bodies, though these would also be unlikely to explain the disc's inner edge. 
Rather, if an as yet unseen outer planet is sculpting this inner edge, it would likely have a semi-major axis close to ${\sim}57$\,au. 
Very massive bodies (even at this radius) can introduce long term linear trends into RV measurements, however no such linear trend is seen by \citet{Marmier13} for q$^1$~Eri. 
We therefore report an upper limit on such a planet's mass of ${<}11\,M_{\rm{Jup}}$ based on the inferred ${\sim}57$\,au semi-major axis, the 13 year measurement baseline, and the radial velocity uncertainty of ${\sim}9\,\rm{ms^{-1}}$, although we note that larger masses may be possible for unfavourable orientations. 

There are two constraints we consider to set a lower-bound on the mass of a planet near the disc inner edge. 
The first is that the planet must be sufficiently massive to truncate the inner edge of the planetesimal belt within the ${\sim}1.4\,\rm{Gyr}$ age of the system. 
This results in a lower limit of $M{>}2\,M_{\oplus}$ \citep[see eq.~3 of][]{Shannon16}. 
The second constraint is that the planet should be more massive than the disc it is carving or else it would be forced to migrate during this interaction \citep{Kirsh09}. 
This results in a stricter lower limit of $M{>}8\,M_{\oplus}$. 

A planet with $M{\sim}8\,M_{\oplus}$ has a secular timescale at the peak brightness radius of the disc that is shorter than the ${\sim}1.4\,\rm{Gyr}$ age of the system \citep[i.e., the timescale over which such a planet would impose structure on the disc, see eq.17 of][]{Pearce14}. 
We may therefore infer that any planet close to the inner edge is unlikely to have an eccentricity significantly greater than that of the inner edge. 
This implies that such a planetary orbit would be close to circular (however we note that if detailed modelling found a better fit with a more realistic eccentric profile, this conclusion would need to be revised). 
We can also infer that such a planet should be currently aligned with the disc mid-plane (or more accurately that secular perturbations would have caused the disc mid-plane to become aligned with the planet), and that the planet should not have started out on an orbit that was inclined to the disc by more than half the scale height (since the alignment process imposes vertical structure on the disc). 
These two constraints could argue against such a planet having been scattered out from a formation location closer to the star due to interactions with other planets (such as q$^1$~Eri-b), since it is unlikely that such scattering would result in a coplanar, eccentric planet. 
However, if such a planet was comparable in mass to the disc, its orbit could be circularised and become aligned with the disc without imprinting significant structure on the disc \citep[e.g.,][]{Pearce15}. 
Alternatively the planet could have formed in situ or migrated out through disc interactions while retaining a nearly circular co-planar orbit.

\subsection{Origin of Disc Asymmetries}
\label{sec:discOriginAsymms}
Our modelling found that the disc's asymmetry is most consistent with a broad clump on the SW ansa with a total flux $1.7{\pm}0.6\,\rm{mJy}$. 
The projected offset of the clump puts it at a radius of ${\sim}80{\pm}15$\,au if it is in the disc mid-plane (i.e., determined from the clump position offsets taking account of the disc position angle and inclination). 
The clump's major and minor axis standard deviations is ${\sim}3.2''\times2.5''$, and its position angle ${\sim}40^{\circ}$ (i.e., angled towards the NW minor axis along the disc inner-edge). 
This is consistent with the observational analysis which inferred a clump located near the inner edge in the SW ansa with a flux of $1.33{\pm}0.29\,\rm{mJy}$ (from the difference between the fluxes in the NE and SW ansae). 
We also found the inner edge to have a radial offset, and so too in the scattered light \textit{HST} imaging. 
In this section we interpret these asymmetric measurements as either due to a disc-planet interaction, a recent collision, or due to extra-galactic background sources.

\subsubsection{A planet-driven asymmetry?}
\label{sec:discPlanAsym}
The discussion in $\S$\ref{sec:discIE} concluded that the inner edge could have been sculpted by a planet. 
Such a planet could have formed closer in and migrated outwards via the exchange of angular momentum with planetesimals in scattering events \citep[i.e., planetesimal-driven migration, see][]{Fernandez84, Ida00, Gomes04, Ormel12}, via interaction with the gas-disc in which it formed, or indeed by interactions with other planets in the system (e.g., with the known $0.93\,M_{\rm{J}}$ planet q$^1$~Eri\,b). 
During this migration its resonances would have swept through the planetesimal disc, and some planetesimals could have become trapped. 
The resonances that become populated in this process would depend on the planet's mass and migration rate and can result in the planetesimal distribution being clumpy \citep{Wyatt03}. 
The clump observed on the inner-edge could therefore be explained by planetesimal trapping in the 2:1 resonance. 
We previously inferred that a planet may have sharpened the inner edge, however if an asymmetry was driven by such a planet's migration, this motion must have stalled as the planet migrated outwards. 
For example, this could occur if the disc ran out of angular momentum, which would imply that the planet mass is comparable with the disc mass, i.e., $M_{\rm{pl}}{\sim}M_{\rm{disc}}{>}8\,M_\oplus$. 
Interesting for this scenario, is that this would be consistent with the evolution of Neptune in our own Solar System, inferred to have stalled during its migration following the resonant sweep-up of planetesimals in the Kuiper Belt \citep{Fernandez84}.

While an eccentric distribution of planetesimals is not needed to explain the sub-mm major axis asymmetry, smaller dust particles are more strongly affected by stellar radiation pressure. 
This additional force places small dust onto eccentric orbits with periastra near to the location of where they were released from their parent planetesimals (presumably in collisions). 
Therefore, small dust created in collisions in the clump would have apastra in the NE of the disc \citep[see][]{Wyatt06}. 
This provides a consistent explanation for the \textit{HST} image which found dust extending to greater distances in the NE (see $\S$\ref{sec:SumObsAnalysis}). 
Thus a planet that arrived at the inner edge of the disc by planet migration could potentially reproduce the observed flux, size and position of the measured asymmetry, in both the sub-mm and scattered light, though further modelling of this scenario is required to investigate this.

Nevertheless, we cannot rule out a scenario in which the asymmetric belt is explained by secular perturbations from an eccentric planet. 
For example, if such a planet migrated towards the inner edge of the belt whilst retaining a sufficiently high eccentricity, this would force an eccentricity into the belt since the planet's secular timescale is less than the age of the system. 
However, since the forced eccentricity of a belt is due to the combination of both the perturbing body's semi-major axis and eccentricity (i.e., to first order, $e_{\rm{forced}}{=} 1.25 e_{\rm{pl}}a_{\rm{pl}}/a_{\rm{p}}$, where $e_{\rm{pl}}$ and $a_{\rm{pl}}$ are the eccentricity and semi-major axis of the planetary perturber, and $e_{\rm{forced}}$ and $a_{\rm{p}}$ are the forced eccentricity and semi-major axis of the perturbed planetesimals), there is a degeneracy between the perturbing body's semi-major axis and its eccentricity. 
Whilst we are therefore unable to tightly constrain these parameters, from the equations 5-8 of \citet{Pearce14}, it can be shown that a planet orbiting at 30\,au with an eccentricity of 0.1 would drive the belt's inner and outer edges to eccentricities of 0.064 and 0.036 respectively (consistent with our derived $R_{\rm{inner}}$ and $R_{\rm{outer}}$ parameters). 
Equivalent eccentricities could be produced from a planet at 60\,au with an eccentricity of just 0.05, i.e., the same semi-major axis as the half-maximum derived inner edge (see Table~\ref{tab:radProfVals}). 
We suggest as further work that such a scenario could be modelled in detail to better constrain required planetary parameters, and to determine if this provides a realistic model for the broad debris disc and its asymmetries.

\subsubsection{A recent collision?}
\label{sec:discCollision}
Alternatively, this clump could have formed from a recent massive collision in the belt \citep[such as those explored in][]{Kral15}. 
This would not require planetesimals to be trapped in resonance (although such a collision could have been at the resonant location). 
Here we consider how large a parent body would need to be for a single collision to reproduce the observed asymmetry. 
By scaling the total dust mass in the disc by the ratio of the $1.33\,\rm{mJy}$ excess flux in the SW ansa to the total disc flux, this implies the clump has a mm dust mass of ${\sim}0.0029\,M_{\oplus}$ (or ${\sim} 30\%$ larger if the $1.7\,\rm{mJy}$ clump flux from model CL had been used). 
For this amount of dust to have formed via a single collision, the parent body must have had a diameter of at least $D{\approx}2300\,\rm{km}$ for a density of $2.7\,\rm{g\,cm}^{-3}$. 
Since this would require the collision to fragment the parent body entirely into mm-sized grains, whereas a range of fragment sizes is more likely, the parent body would most likely be significantly larger than 2300\,km.

While \S \ref{sec:discMASS} concluded that no bodies larger than $D_{\rm{max}}{\approx}1$\,km need to be present in the disc, if the $\alpha{=}3.5$ \citet{Dohnanyi69} size distribution continues from $D_{\rm{max}}$ to 2300\,km then the collisional lifetime of 2300\,km bodies can be estimated to be a factor of $\sqrt{2300\,\rm{km}/D_{\rm{max}}}$ longer than the age of the system, i.e., ${\sim}40$\,Gyr. 
Extrapolating from the number of $2\,\rm{cm}$ bodies present in the $0.028\,M_{\oplus}$ disc, an $\alpha{=}3.5$ size distribution implies that there would be ${\sim} 10^{4}$ bodies of size ${\sim} 2300$\,km present in the disc, and therefore we would expect collisions to occur every ${\sim}4\,\rm{Myr}$. 
Clumps disperse after formation, and Fig.7 of \citet{Jackson14} suggests that this process might take ${\sim} 1000$ orbits, i.e., ${\sim} 0.5$\,Myr at ${\sim}60$\,au. 
Comparing this dispersal rate to the expected collision rate, this calculation suggests that although infrequent, clumps are not implausibly rare in this disc, demonstrating a ${\sim}10\%$ probability that we might observe a clump in the disc at any one time if such massive planetesimals exist in the disc and collisions are energetic enough to fragment these. 

However, there are two arguments against this. 
Most significantly, collisions between even 2300\,km-sized bodies may not be sufficient to cause the observed clump, since these would not be 100\% efficient at converting parent bodies into mm-sized dust, i.e., we would very likely need even larger collisional bodies. 
The second is that the disc mass required to get even a $10\%$ detection fraction in this scenario is ${\sim}300\,M_{\oplus}$, which as discussed in $\S$\ref{sec:discvertscal}, may be problematic. 
Such a high mass is inevitable if there are to be enough 2300\,km bodies for collisions to occur frequently, although we note here that the presence of such large bodies might help to explain the observed scale height of the disc. 
If we were to instead assume the size distribution was flatter, then we could raise the clump detection probability, however this would require an even higher disc mass. 
For example, to raise the detection probability to order unity, there would instead need to be $10^{5}$ ${\sim} 2300$\,km bodies, however such a disc would then have a mass ${\gtrsim}3000\,M_{\oplus}$. 
This scenario would then be problematic for the reason outlined in $\S$\ref{sec:discvertscal}, i.e., the disc mass would then exceed the solid mass available to form planetesimals in protoplanetary discs. 
Therefore, whilst a collisional origin for this asymmetry could explain the observed asymmetry in both the sub-mm and scattered light, it is perhaps unlikely.

\subsubsection{Clumpiness from extra-galactic emission?} 
\label{sec:discSMG}
An extra-galactic sub-mm galaxy (SMG) in the SW ansa is one possible explanation for the asymmetry seen in the \textit{ALMA} images. 
Perhaps the strongest argument against this is that such an SMG would not explain the extended scattered light emission in the NE, and the simplest explanation for the asymmetries measured in the sub-mm and in scattered light is that they are caused by a single phenomenon. 
Nevertheless, a more complicated scenario involving two phenomena cannot be ruled out. 
Thus, we also consider the probability of detecting an SMG coincident with the disc. 
From the SMG counts of \citet{Simpson15}, we find a detection probability (within a projected $200$\,au from the image centre with $\delta F{\sim}1.33\,\rm{mJy}$) of $P(\rm{SMG}){\sim}0.2\%$. 
Whilst not implausibly rare, this casts doubt on this interpretation. 
We also found this clump to be resolved or azimuthally broadened, which would require this SMG to either be an extended source, or be due to multiple sources nearby on the sky. 
However, the frequency of SMGs appearing nearby around this flux level, and the number of large SMGs (i.e., $>0.7''$, see $\S$\ref{sec:obsInts}) are both rare, additionally arguing against an SMG interpretation.

The spectral index has also been used to discriminate between possible origins of sub-mm emission \citep{Su17, Booth19}, since this differs between optically thin dust from planetesimal collisions, $\alpha_{\rm{mm}}{\approx}2{-}2.5$, and extra-galactic emission, $\alpha_{\rm{mm}}{\approx}3{-}4$. 
In $\S$\ref{sec:FPAObsAn} we derived the spectral index for the full disc of $\alpha_{\rm{mm}} {=} 2.34{\pm}0.29$, which is consistent with measurements of other debris discs \citet{Ricci15}. 
By instead considering the spectral index due to the flux difference between the SW and NE ansae, $\delta F$, in Bands 6 and 7 (see Table~\ref{tab:fluxVals}), we find only a weak constraint on the spectral index of the clump $\alpha_{\rm{mm}}{\sim}9{\pm}4$ (i.e., consistent with the spectral index for thin dust or an SMG). 
We therefore cannot claim the spectral index of q$^1$~Eri as inconsistent with either optically thin dust or extra-galactic emission, either from its total sub-mm fluxes, or those of the clump. 
Whilst we cannot exclude the possibility that the clump is due to extra-galactic emission, this interpretation seems less likely than a planet-driven asymmetry (see $\S$\ref{sec:discPlanAsym}) since this cannot simultaneously explain the scattered light asymmetry and such an SMG would have a very low occurrence rate. 
Further \textit{ALMA} observations could ascertain if this inner edge asymmetry is co-moving with the belt in the future and definitively rule this out as due to background SMG emission based on the signal-to-noise of the clump. 
For example, if this can be measured with a SNR${>}3$ with the same beam in 2018, given the proper motion of the system is ${\sim}200\,\rm{mas}\,\rm{yr}^{-1}$, and the precision to which a Gaussian can be centred on this emission is $0.5\times \rm{beamsize} / \rm{SNR}$, measuring a ${>}3\,\sigma$ change in this location will require waiting a further ${\sim}2\,\rm{yr}$ from the 2018 observations (i.e., such a change may now be measurable). 

\subsubsection{Interactions with the ISM?}
Although interactions with the interstellar medium (ISM) are not expected to affect the orbits of the sub-mm grains, the ISM can significantly influence the orbits of smaller micron-sized dust observed in the scattered light. 
Indeed, this has been shown to be the case with HD\,15115, which shows a swept-back asymmetric disc in scattered light as observed with \textit{HST} \citep[for which the East-West emission shows a radial offset of 2, see][]{Kalas07}, however when observed with \textit{ALMA}, there are no observed radial offset asymmetries \citep[see][]{MacGregor19}. 
This suggests that there does not to be a single mechanism to interpret sub-mm and scattered light asymmetries simultaneously. 
Whilst a single mechanism may still however be preferable for probabilistic reasons, we cannot rule out the possibility that the asymmetries observed in the scattered light and the sub-mm are independent, i.e., with both ISM interactions and planet-disc interactions affecting these respectively.

\subsection{What is the Inner Warm Component?}
\label{sec:discussionWarm}
q$^1$~Eri has previously been found to be consistent with having a multi-component debris disc, comprising an outer cool belt with $r_{\rm{bb}}{\sim}60$\,au, and an inner warm belt with $r_{\rm{bb}}{\sim}10$\,au \citep{Kennedy14}. 
\citet{Schuppler16} likewise modelled this system, and found two disc components, an inner warm belt between $3{-}10$\,au, close to the known exo-Jupiter at ${\sim}2$\,au, and an outer cool belt between $75{-}125$\,au. 
Although our \textit{ALMA} images clearly resolved the disc's outer belt, the only other significant Band~7 emission detected was unresolved and coincident with the star with a flux of $169{\pm}22\,\mu\rm{Jy}$. 
Given that our flux distribution model predicts the $856\,\mu$m stellar photospheric flux to be ${\sim}99\,\mu\rm{Jy}$ (with a 2\% uncertainty), this suggests that at the ${\sim}3\,\sigma$ level, we may have detected additional emission coincident with the location of the star with a flux $F_{\rm{inner}}{=}70{\pm}22\,\mu$Jy. 
This is consistent with what might be expected for the emission of an inner warm component (see Fig.~\ref{fig:SED}), however we note that the uncertainties associated with this modelled inner component are large since the slope of the inner modified blackbody are determined entirely by the two \textit{ALMA} Band~6 and Band~7 data points (i.e., there are no resolved measurements of the inner component in the near, mid or far-infrared). 
Since the angular size of emission at ${\sim}10$\,au from the star at a distance of $17.34$\,pc would be ${\sim}0.6\arcsec$, i.e., comparable with our beam size, emission internal to this radius would be largely unresolvable from the stellar emission in our images. 
This emission could arise from a planetesimal belt inside 10\,au, be due to additional stellar emission, or image noise coincident with the stellar position. 
We discuss each of these in turn. 

We first consider whether this emission is evidence of a warm planetesimal belt internal to 10\,au, which if present, may resemble the Solar System's asteroid belt, though with higher mass and external to the orbit of the known exo-Jupiter, q$^1$~Eri\,b. 
Belts consistent with this have been previously inferred towards q$^1$~Eri by \citet{Kennedy14} with a blackbody radius $r_{\rm{bb}}=$10\,au (and consistent with the flux distribution presented here in Fig.~\ref{fig:SED},) and at 3\,au by the flux distribution modelling of \citet{Schuppler16} from the mid-IR data.
Even though this blackbody radius is relatively well constrained, such an assessment does not come without uncertainty. 
The value determined by \citet{Kennedy14} is dominated by emission from the mid-IR flux measurements (i.e., not the longer wavelength \textit{ALMA} data), and such radius estimations can significantly underestimate the true radii of discs \citep[see equation 8 of][]{Pawellek15}. 
For example, for a $50\%$ astrosilicate grain-$50\%$ ice dust composition around a star with a luminosity of $L{=}1.59\,L_{\odot}$ (consistent with q$^1$~Eri, see $\S$\ref{sec:SED}), the true belt radius could be larger than the blackbody radius by up to a factor of ${\sim}4.6$. 
If so, the warm emission analysed by \citet{Kennedy14} could have a contribution from dust as far out as ${\sim}50$\,au, i.e., dust on the inner edge of the disc, and so be potentially associated with the clump rather than an inner planetesimal belt. 

However, for two reasons we believe this to be highly unlikely.
Firstly, we used the IRS data to consider the possibility of the warm
emission being associated with the inner edge clump at ${\sim}60$\,au ($3.5\arcsec$). 
This clump would lie within the IRS slit for the ``long" modules 
($\lambda{>}$14\,$\mu$m), and could feasibly shift the centroid of IRS emission as it changes from star to disc-dominated (as seen in the flux distribution in Fig.~\ref{fig:SED}). 
However, the spatial profiles and PSF-subtracted residual images in the CASSIS database \citep{Lebouteiller11} appear consistent with a single unresolved source across all wavelengths, so it is unlikely that any clump has contributed flux to the IRS spectrum. 
Secondly, we note that applying the same correction factor to the main belt would lead us to predict its belt to be at ${\sim}276$\,au, whereas it is observed at $81.6$\,au, a factor of just ${\sim}1.4$ larger than $r_{\rm{bb}}$.
Although the inner and outer belt correction factors may be independent (e.g., if they are compositionally different), this suggests that such a high correction factor may still be unrealistic. 
If applied accordingly, such a correction factor would predict the inner planetesimal belt to have a radius of ${\sim}14$\,au, and (at the flux observed) be difficult to resolve from the star with this \textit{ALMA} data. 
Therefore, since this emission seems unlikely to have been confused with the outer planetesimal belt, we cannot rule out the presence of a high mass planetesimal belt internal to 10\,au, which like in \citet{Marino18}, we have shown may be detectable in the sub-mm.

If such a belt was present, and approximated a narrow Gaussian ring, equation 5 of \citet{Matra20} predicts that in visibility space this would produce a Bessel function with a first null at ${\sim}130\,$k$\lambda$. 
This might suggest that such a function may be visible near such baselines (for example, in Fig.~\ref{fig:VisBase}). 
However on inspecting this visibility data, given the SNR of emission at these longest baselines (i.e., shortest angular scales) no such function is visible with significant emission at the ${\sim}70\,\mu$Jy level, even when the bin sizes are increased to raise the SNR per bin. 
Given this warm emission is faint, this analysis therefore cannot exclude the existence of a narrow inner belt inside 10\,au.

One alternative to explain this emission could be that it has originated in sub-mm stellar flaring events.
Although common around M-type stars \citep[see][]{MacGregor20}, second-minute timescale sub-mm variability can be detected towards solar-type main-sequence stars at the level of tens-hundreds of $\mu$Jy, consistent with the warm excess measured here \citep{Burton21}. 
Although between the two Band~7 epochs (separated by two years) we found consistent flux measurements (within 10\,$\mu$Jy) at the location of the star, we did not explore stellar variability in detail in this analysis.
Therefore, we equally cannot rule out the possibility that this warm excess is due to stellar variability from sub-mm flares. 

Finally, we note that we cannot rule out this emission as due to noise coincident with the stellar location, although at the $3\,\sigma$ level, or contamination with the disc flux.
Although this appears less plausible than the other two interpretations, if the inner edge of emission on the minor axis is only marginally resolved from the star, then this could have contributed to the measured excess. 
Therefore, to better understand the nature of this inner component, we note that further high-resolution scattered light, mid-IR and sub-mm measurements (e.g., with \textit{SPHERE}, \textit{JWST} and higher resolution and deeper \textit{ALMA} imaging) are necessary.

\begin{figure}
    \centering
    \includegraphics[width=1.0\columnwidth]{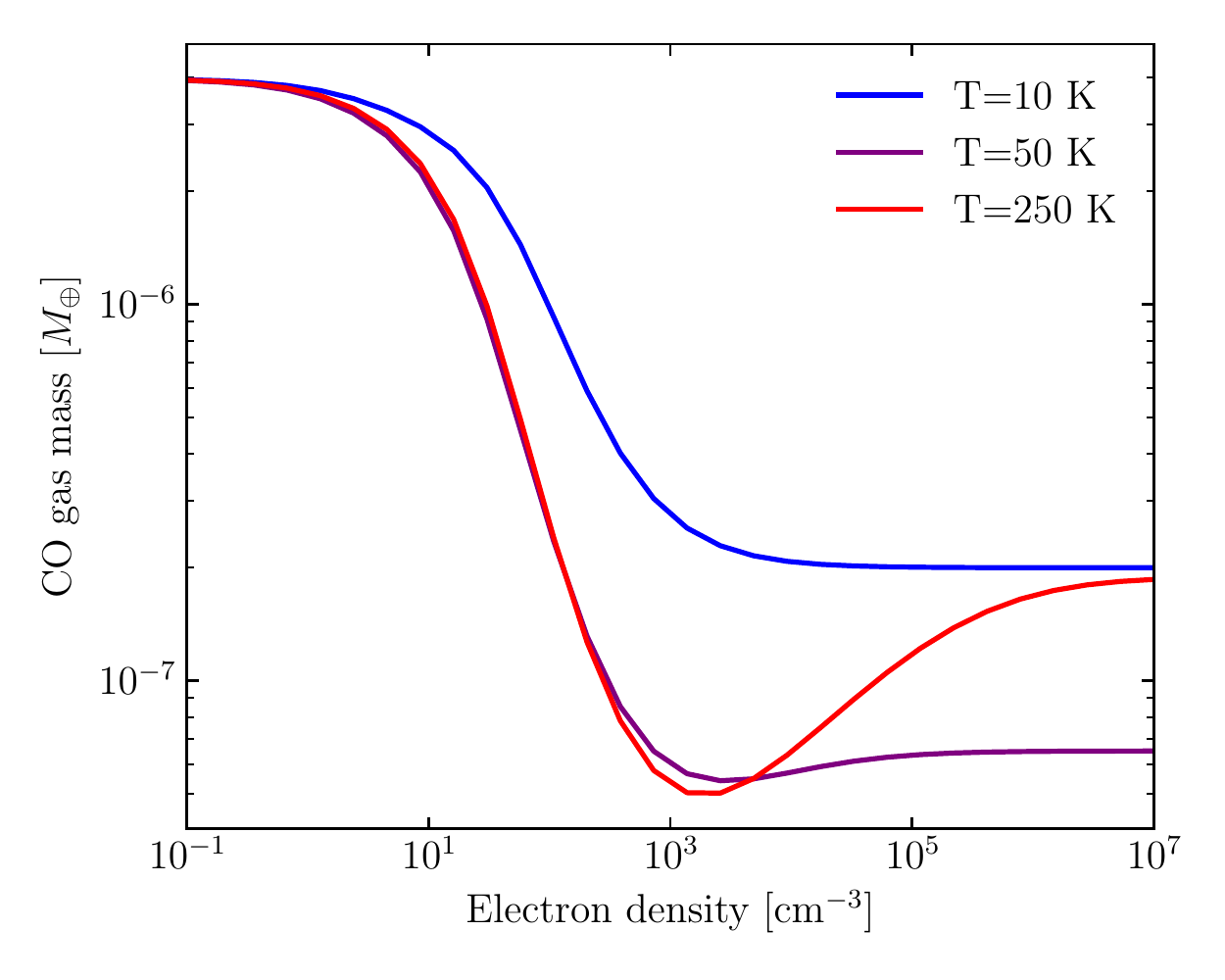}
    \caption{CO gas mass as a function of the electron density, for the three temperatures of 10K, 50K and 250K.}
    \label{fig:coGasMass}
\end{figure}

\subsection{CO Gas Mass}
\label{sec:discussionCO}
In $\S$\ref{sec:discCO} we demonstrated that $\rm{CO}$ spectral signatures for the J=3-2 transition line are not present in our \textit{ALMA} data.  
This non-detection of $\rm{CO}$ resulted in an upper limit on the $\rm{CO}$ flux of $24.0\,\rm{mJy\,kms^{-1}}$. 
This can be used to derive an upper limit on the gas mass from the excitation conditions of the gas set by the radiation environment, electron density and kinetic temperature \citep{Matra15}. 
We compute the level populations of the CO J=3 rotational level using an NLTE (Non-Local Thermodynamic Equilibirum) code including fluorescence \citep{Matra15, Matra18}. 
For three temperatures $10\,\rm{K}$, $50\,\rm{K}$ and $250\,\rm{K}$ (covering a range of values appropriate for debris discs), together with the stellar flux (see Fig.~\ref{fig:SED}) and the peak emission radius of ${\sim}81.6$\,au, Fig.~\ref{fig:coGasMass} shows the estimated $\rm{CO}$ gas mass as a function of the electron density in the disc. 
We plot a range of collision-partner densities, from low densities where molecular excitation is dominated by radiation (e.g., fluorescence, from electron excitation by starlight, followed by decay through higher J rotational levels), through to larger densities where line excitation is instead dominated by collisions, and as such the line populations are in local thermodynamic equilibrium where the line fluxes depend only on temperature. 
From this we can then set an NLTE upper bound CO gas mass of $M_{\rm{CO}} {<} 4 {\times} 10^{-6}\,M_{\oplus}$, the constraint for low collider densities.

Based on the assumption that icy planetesimals create second-generation CO gas through planetesimal collisions, \citet{Kral17} predicted a $\rm{CO}$ gas mass in q$^1$~Eri's disc to be $1.1 \times 10^{-7}M_{\oplus}$, based on the predicted collision rate of bodies from mid and far-IR flux measurements, a stellar luminosity and temperature consistent with values in this work, and a disc radius of $105$\,au. 
Such a predicted gas mass is consistent with the upper limit derived here, being two orders of magnitude lower than the upper bound mass given above. 
From equation 2 of \citet{Matra17} we can use this upper limit $\rm{CO}$ mass alongside the stellar mass, luminosity, disc fractional luminosity, disc peak emission radius and belt width, with a CO photo-dissociation time of 120 years \citep[see][]{Visser09} to estimate an upper bound on the fraction of planetesimals composed of $\rm{CO}+\rm{CO_{2}}$, which is $f_{\rm{CO+CO_{2}}}<95\%$. 
This upper bound is consistent with the Solar System ($f_{\rm{CO+CO_{2}}}\sim10\%$). 
Nevertheless, higher sensitivity \textit{ALMA} data could place tighter constraints on this upper limit CO gas mass, and if measured with sufficient depth, detect any CO, if present. 
Furthermore, the presence of different chemical species can be constrained by ALMA (for example, CI, OH, HCN, N$_2$H$+$). 
With measurements of such other gaseous species, in conjunction these may provide us with a far better understanding of the composition of this debris disc.

\section{Conclusions}
\label{sec:conclusions}
We have presented Band~6 and 7 \textit{ALMA} observations, and \textit{HST} scattered light observations of the q$^1$~Eri debris disc.
Until now the debris disc architecture has been interpreted from low resolution thermal imaging (${\sim}4 \arcsec$) and scattered light HST imaging with a poor inner working angle (${\sim}2.25\arcsec$). 
Here we have explored the full extent of the q$^1$~Eri debris disc at sub-arcsecond resolution, placing bounds on disc features such as the inner and outer edges, peak emission radius, extent, inclination, position angle, the vertical dust distribution, and placed an upper limit on the CO gas mass and on the ice mass fraction of $\rm{CO}+\rm{CO_2}$ present. 

From the 1.4\,Gyr age of the system, we placed lower limits on the size of the largest planetesimals in this disc from their collisional lifetimes, finding $D_{\max}{>}1\,\rm{km}$, leading to an estimate of the mass of this disc of $M_{\rm{disc}}{>}8M_{\oplus}$, suggesting that whilst this disc is an outlier in brightness, it need not be an outlier in terms of its mass. 
From observations and modelling that tentatively determined this disc to be resolved in its vertical direction, we investigated the size and mass of bodies that could be responsible for stirring the disc. 
This interpretation required bodies with a size at least $1200$\,km significantly above the lower limit on the largest planetesimal size, although the disc could alternatively have been born stirred (or planet-stirred), in which case such large planetesimals would not be necessary. 

We have demonstrated through image analysis and modelling that a series of asymmetric features exist in the sub-mm emission: with a total flux asymmetry between the SW and NE sides, and a radial offset towards the SW ansa in the major axis. 
Our modelling found this disc is most consistent with an axisymmetric disc with an emission clump on its SW inner edge. 

At just ${\sim}2$\,au, the known planet  q$^1$~Eri\,b is too close in to affect the observed outer main belt, and we discuss how the broad disc may be connected to other aspects of this planetary system, such as other potential planets and planetesimal collisions. 
One interpretation of this system is that the belt of planetesimals had its inner edge carved by a planet which formed a clump on the inner edge whilst migrating, similar to the inferred evolution of Neptune. 
This scenario may simultaneously explain the inner edge sharpness, the SW ansa sub-mm clump and the scattered light asymmetry on the outer edge in the NE. 

By assessing the emission coincident with the star, and by comparing this with previous modelling and our flux distribution, we show that there is tentative evidence for an inner warm component. 
This may be due to a belt of planetesimals closer to the planet q$^1$~Eri~b, between radii $3{-}10$\,au, and thus a high mass Asteroid Belt analog. 

Throughout this work we have suggested further analysis that could be undertaken to better understand this disc. New discoveries and confirmations of interpretations laid out here could be made with more detailed modelling, and future observations with instruments/observatories such as \textit{ALMA}, \textit{HST}, \textit{SPHERE} and \textit{JWST}.

\section*{Data Availability Statement}
This work makes use of the following \textit{ALMA} data: ADS/JAO.ALMA\ 2017.1.00167.S, 2015.1.01260.S, and 2015.1.00307.S. 
\textit{ALMA} is a partnership of ESO (representing its member states), NSF (USA) and NINS (Japan), together with NRC (Canada), MOST and ASIAA (Taiwan), and KASI (Republic of Korea), in cooperation with the Republic of Chile. 
The Joint \textit{ALMA} Observatory is operated by ESO, AUI/NRAO and NAOJ. 
Based on observations made with the NASA/ESA Hubble Space Telescope, and obtained from the Hubble Legacy Archive, which is a collaboration between the Space Telescope Science Institute (STScI/NASA), the Space Telescope European Coordinating Facility (ST-ECF/ESA) and the Canadian Astronomy Data Centre (CADC/NRC/CSA), program 10539. 
This work has made use of data from the European Space Agency (ESA) mission {\it Gaia}\footnote{\url{https://www.cosmos.esa.int/gaia}}, processed by the {\it Gaia} Data Processing and Analysis Consortium (DPAC\footnote{\url{https://www.cosmos.esa.int/web/gaia/dpac/consortium}}).
Funding for the DPAC has been provided by national institutions, in particular the institutions participating in the {\it Gaia} Multilateral Agreement. 
This research has made use of the NASA Exoplanet Archive, which is operated by the California Institute of Technology, under contract with the National Aeronautics and Space Administration under the Exoplanet Exploration Program.
The Combined Atlas of Sources with Spitzer/IRS Spectra (CASSIS) is a product of the Infrared Science Center at Cornell University, supported by NASA and JPL. 
This work has made use of IRS data with a programme ID: 20463.

\section*{Acknowledgements}
We thank the anonymous reviewer for their comments which improved the quality of this work.
JBL is supported by an STFC postgraduate studentship. 
SM is supported by a Research Fellowship from Jesus College, Cambridge.
GMK is supported by the Royal Society as a Royal Society University Research Fellow. 
The research of OP is funded through the Royal Society Dorothy Hodgkin Fellowship. 
TDP is supported by DFG grants Kr 2164/14-2 and Kr 2164/15-2.


\bibliographystyle{mnras}
\bibliography{example} 


\appendix
\section{Visibility Data}
\label{sec:appendixVis}
\begin{figure*}
    \includegraphics[width=2.0\columnwidth]{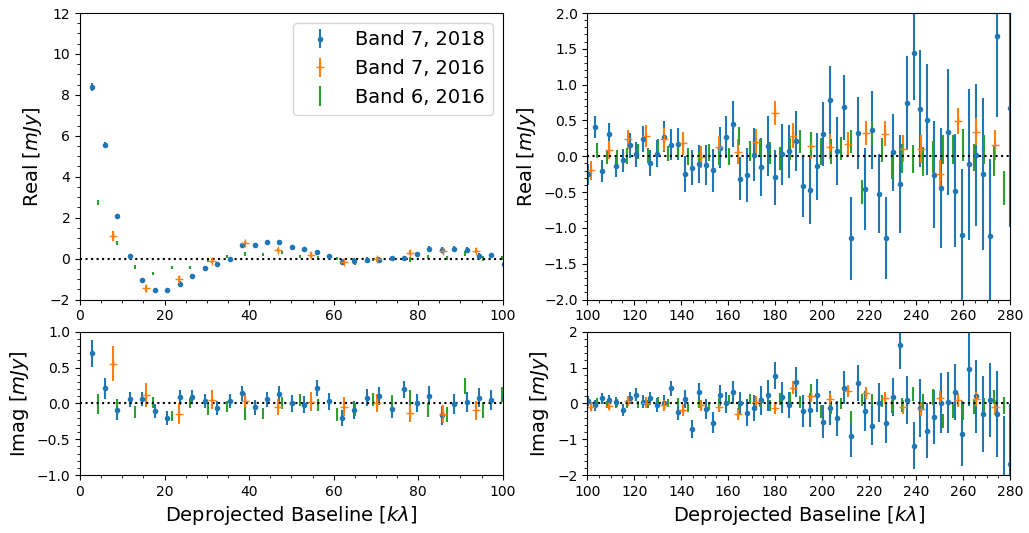}
    \caption{Composite plots of the binned visibility data for the three \textit{ALMA} epochs. Error bars here are calculated from the standard deviation divided by the number of independent data points within each bin, where we have included 120 bins per measurement set.}
    \label{fig:VisBase}
\end{figure*}

For completeness, we include here the visibility data for all three \textit{ALMA} epochs. 
Using the derived values for $i$ and $\rm{PA}$, Fig.~\ref{fig:VisBase} shows the interferometric visibilities as a function of the deprojected baselines. 
Although these profiles differ (their absolute values of their maxima and minima are inconsistent), they show consistently located nulls at ${\sim}10\,\rm{k}\lambda$, ${\sim}35\,\rm{k}\lambda$ and ${\sim}60\,\rm{k}\lambda$, and consistently located peaks at ${\sim}20\,\rm{k}\lambda$ and ${\sim}45\,\rm{k}\lambda$. 
This suggests that although these data sets may have different profiles, their radial emission has brightness peaks and minima at similar radial locations. 
The imaginary data (seen in the lower plots, which probe the azimuthal structure) shows a number of significant features (particularly at the largest angular scales with uv-distances below $10\,\rm{k}\lambda$ for the 2018 Band~7 data). 
Although the majority of imaginary visibility data points are consistent with zero (i.e., axisymmetry), for the very shortest baselines (i.e., those associated with larger scale structure) there are departures from this, symptomatic of the type of departure from symmetry as measured in $\S$\ref{sec:radialProfAn}.

\section{Dust Density and Temperature Distribution}
\label{sec:appendixDust}
\begin{figure*}
    \centering
    \includegraphics[width=2.0\columnwidth]{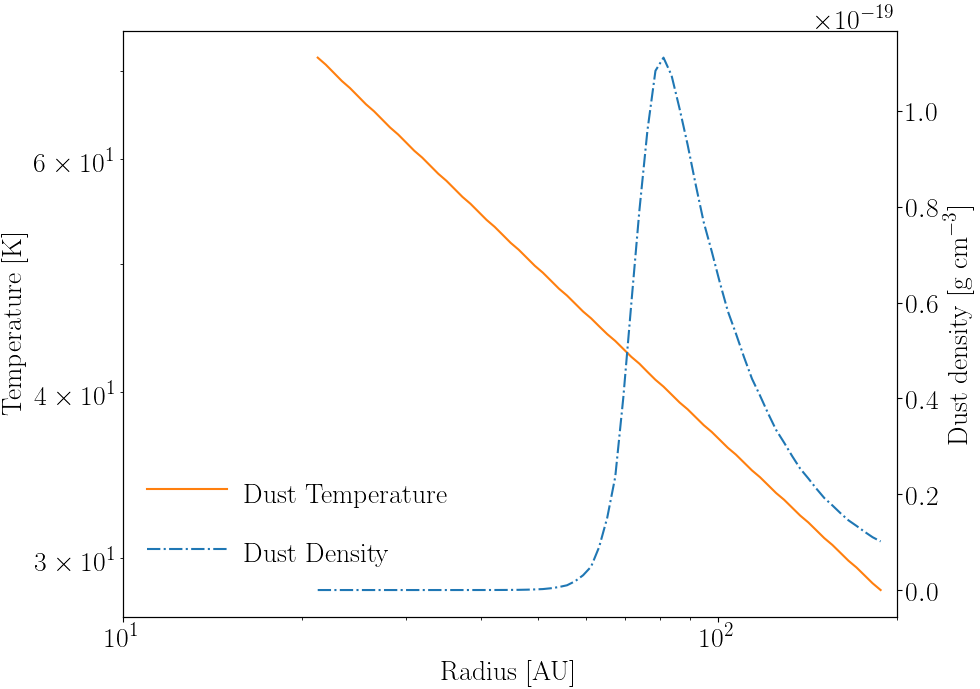}
    \caption{Dust temperature and dust density as a function of radial distance from the star, as computed by $\rm{RADMC-3D}$. }
    \label{fig:DustTempDen}
\end{figure*}

Here we provide the dust and temperature distributions as calculated by $\rm{RADMC-3D}$ for the best-fit model \textbf{CL} in Fig.\ref{fig:DustTempDen}. 

\section{Posterior Distribution Outputs}
\label{sec:appendixA}
\begin{figure*}
    \centering
    \includegraphics[width=2.0\columnwidth]{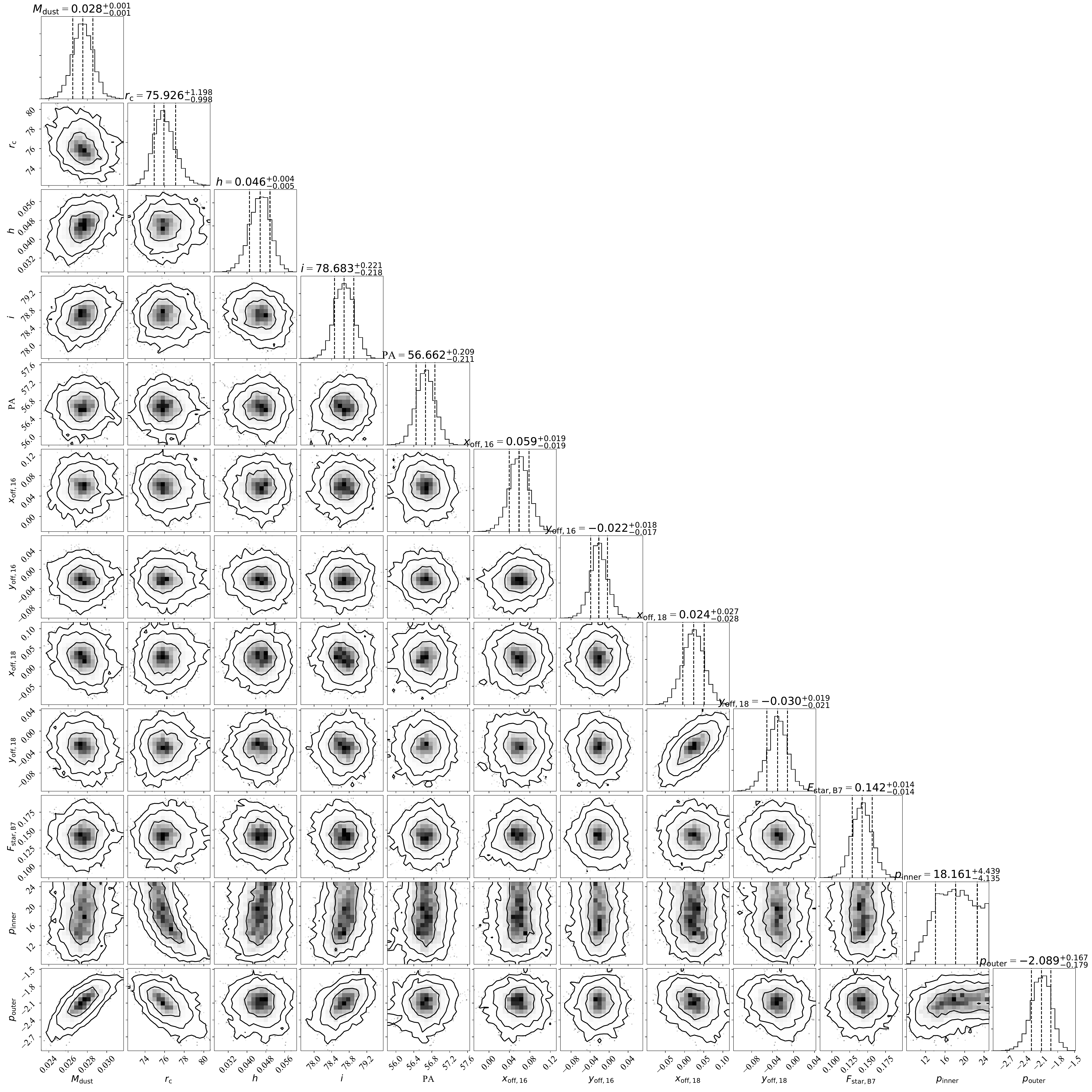}
    \caption{MCMC posterior distribution of the 18 parameter ``Model CL": the symmetric belt with clump in the SW ansa. Posterior distributions show the range of values determined for each parameter (see each histogram), and how each parameter varies as a function of each of the other parameters. This allows us to interpret whether the fits are well constrained, and if there are any degeneracies between variables. Here we show the 12 parameters which describe the underlying disc. It can be seen that the distribution of the parameter $p_{\rm{inner}}$ is flat beyond a value of 18, meaning that this is poorly constrained.}
    \label{fig:appCornerPlot1}
\end{figure*}

\begin{figure*}
    \centering
    \includegraphics[width=2.0\columnwidth]{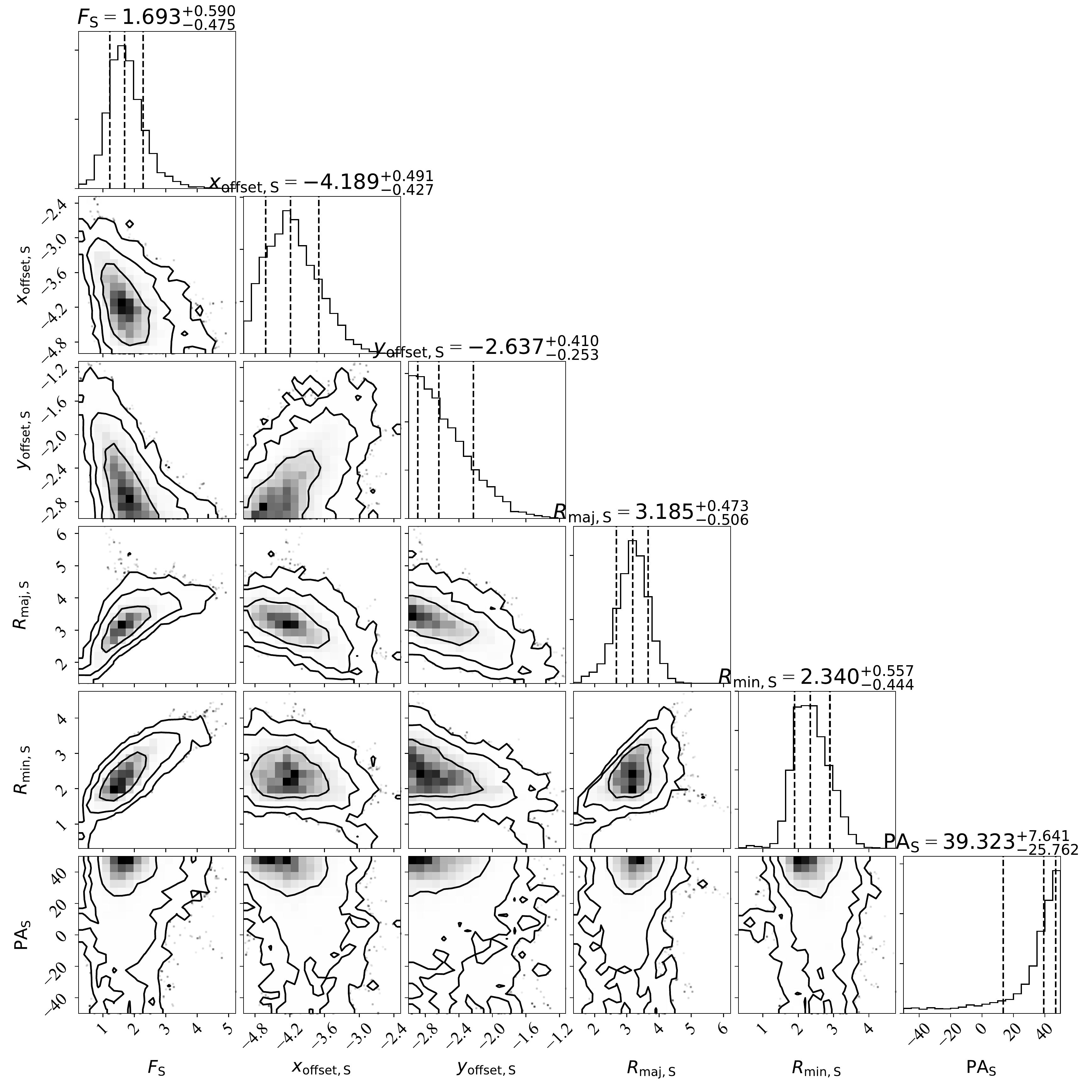}
    \caption{MCMC posterior distribution, as per Fig.~\ref{fig:appCornerPlot1}. Here we show the 6 parameters which describe the SW clump. }
    \label{fig:appCornerPlot2}
\end{figure*}

Here we provide a visual of the MCMC posterior distribution corner-plots for the model with a clump in the SW ansa (model CL), in two plots showing the disc parameters (in Fig.~\ref{fig:appCornerPlot1}) and for the SW clump (in Fig.~\ref{fig:appCornerPlot2}).

\bsp	
\label{lastpage}
\end{document}